%% file: main.tex
\documentclass[twocolumn,astrosymb,twocolappendix]{aastex7}
\usepackage{xspace}
\usepackage{CJKutf8}
\usepackage{bm}
\usepackage{appendix}
\usepackage{amsmath,amssymb}

\newcommand{\sbunit}{\mathrm{mag\ arcsec}^{-2}}

\newcommand{\elvesdwarf}{\textsc{ELVES-Dwarf}\xspace}

\newcommand{\code}[1]{\textbf{\texttt{#1}}}
\newcommand{\sersic}{S\'ersic\xspace}
\newcommand{\satgen}{\textsc{SatGen}\xspace}
\newcommand{\kms}{\mathrm{km\ s^{-1}}}
\newcommand{\rvir}{$R_{\rm vir}$\xspace}
\newcommand{\nsat}{$N_{\rm sat}$\xspace}

\usepackage{xcolor}
\usepackage{enumitem}

\defcitealias{Greco2018}{G18}
\defcitealias{Karachentsev2013}{K13}
\defcitealias{McConnachie2012}{Mc12}
\defcitealias{Carlsten2021}{C21}
\defcitealias{Tully2013}{T13}
\defcitealias{Tully2016}{T16}
\defcitealias{Fingerhut2010}{F10}
\defcitealias{McQuinn2017}{M17}
\defcitealias{Manwadkar2022}{M22}
\defcitealias{Danieli2023}{D23}

\shorttitle{ELVES-Dwarf Survey I}
\shortauthors{Li et al.}
\graphicspath{{./}{figures/}}
\begin{document}
\begin{CJK*}{UTF8}{gbsn}

\title{\elvesdwarf I: Satellites Systems of Eight Isolated Dwarf Galaxies in the Local Volume}

\correspondingauthor{Jiaxuan Li}

\author[0000-0001-9592-4190]{Jiaxuan Li (李嘉轩)}
\email[show]{jiaxuanl@princeton.edu}
\affiliation{Department of Astrophysical Sciences, 4 Ivy Lane, Princeton University, Princeton, NJ 08540, USA}
\author[0000-0002-5612-3427]{Jenny E. Greene}
\email{jgreene@astro.princeton.edu}
\affiliation{Department of Astrophysical Sciences, 4 Ivy Lane, Princeton University, Princeton, NJ 08540, USA}

\author[0000-0002-1841-2252]{Shany Danieli}
\email{sdanieli@tauex.tau.ac.il}
\affiliation{Department of Astrophysical Sciences, 4 Ivy Lane, Princeton University, Princeton, NJ 08540, USA}
\affiliation{School of Physics and Astronomy, Tel Aviv University, Tel Aviv 69978, Israel}

\author[0000-0002-5382-2898]{Scott G. Carlsten}
\email{scarlsten@gmail.com}
\affiliation{Department of Astrophysical Sciences, 4 Ivy Lane, Princeton University, Princeton, NJ 08540, USA}

\author[0000-0002-7007-9725]{Marla Geha}
\email{marla.geha@yale.edu}
\affiliation{Department of Astronomy, Yale University, New Haven, CT 06520, USA}

\author[0000-0001-6115-0633]{Fangzhou Jiang}
\email{fangzhou.jiang@pku.edu.cn}
\affiliation{Kavli Institute for Astronomy and Astrophysics, Peking University, Beijing 100871, China}

\author[0000-0002-5011-5178]{Masayuki Tanaka}
\email{masayuki.tanaka@nao.ac.jp}
\affiliation{National Astronomical Observatory of Japan, Osawa 2-21-1, Mitaka, Tokyo 181-8588, Japan}

%% Note that the \and command from previous versions of AASTeX is now
%% depreciated in this version as it is no longer necessary. AASTeX 
%% automatically takes care of all commas and "and"s between authors names.

%% AASTeX 6.31 has the new \collaboration and \nocollaboration commands to
%% provide the collaboration status of a group of authors. These commands 
%% can be used either before or after the list of corresponding authors. The
%% argument for \collaboration is the collaboration identifier. Authors are
%% encouraged to surround collaboration identifiers with ()s. The 
%% \nocollaboration command takes no argument and exists to indicate that
%% the nearby authors are not part of surrounding collaborations.

%% Mark off the abstract in the ``abstract'' environment. 
\begin{abstract}
The satellite populations of Milky Way--mass systems have been extensively studied, significantly advancing our understanding of galaxy formation and dark matter physics. In contrast, the satellites of lower-mass dwarf galaxies remain largely unexplored, despite hierarchical structure formation predicting that dwarf galaxies should host their own satellites. We present the first results of the \elvesdwarf survey, which aims to statistically characterize the satellite populations of isolated dwarf galaxies in the Local Volume ($4<D<10$~Mpc). We identify satellite candidates in integrated light using the Legacy Surveys data and are complete down to $M_g\approx -9$~mag. We then confirm the association of satellite candidates with host galaxies using surface brightness fluctuation distances measured from the Hyper Suprime-Cam data. We surveyed 8 isolated dwarf galaxies with stellar masses ranging from sub-Small Magellanic Cloud to Large Magellanic Cloud scales ($10^{7.8} < M_\star^{\rm host}<10^{9.5}\, M_\odot$) and confirmed 6 satellites with stellar masses between $10^{5.6}$ and $10^{8} \, M_\odot$. Most confirmed satellites are star-forming, contrasting with the primarily quiescent satellites observed around Milky Way--mass hosts. By comparing observed satellite abundances and stellar mass functions with theoretical predictions, we find no evidence of a ``missing satellite problem'' in the dwarf galaxy regime.
\end{abstract}


\section{Introduction}\label{sec:intro}

\begin{figure*}
    \centering
    \includegraphics[width=1\linewidth]{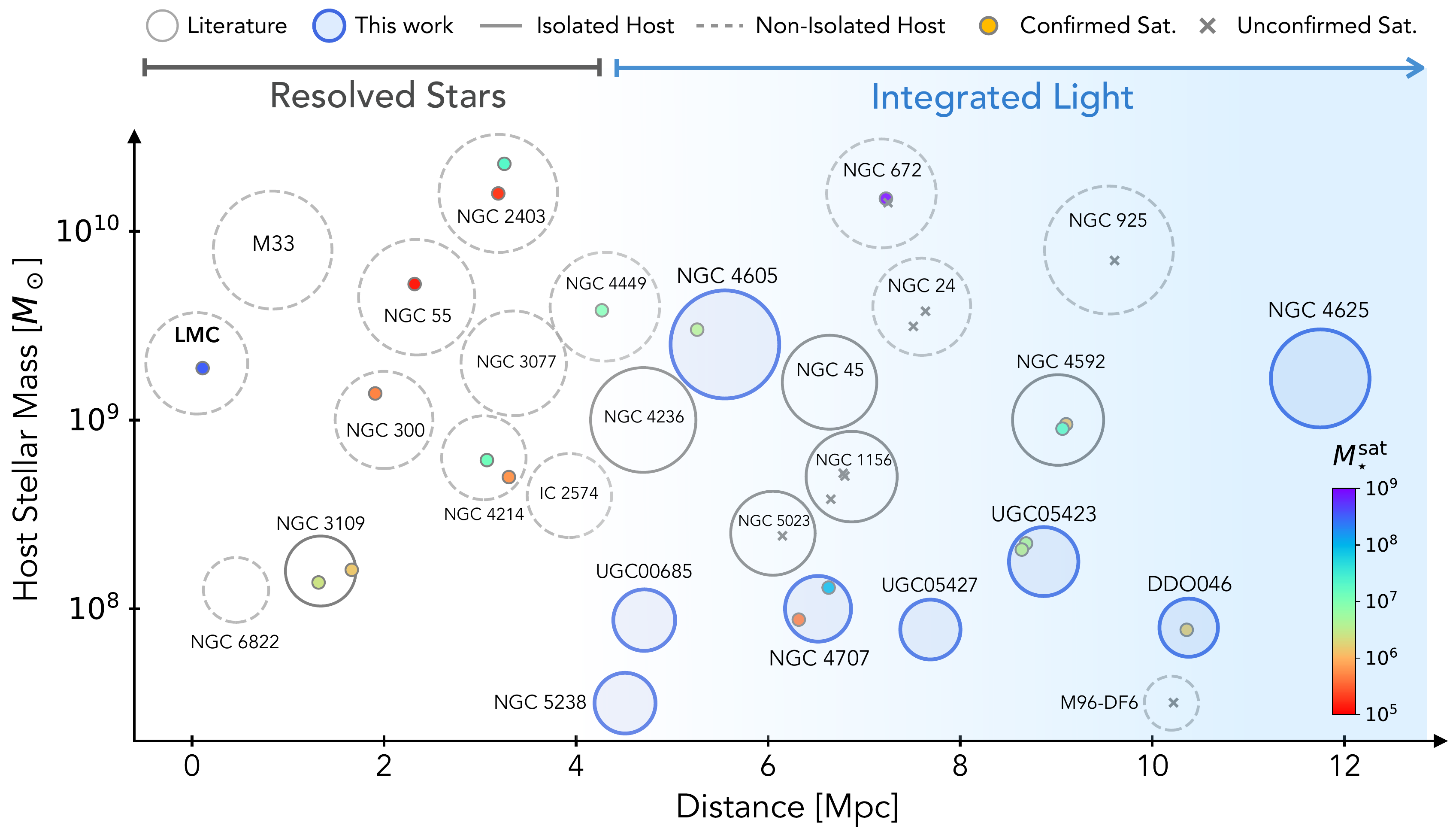}
    \caption{
    Demography of satellites of dwarf galaxies in the Local Volume, arranged by host stellar mass and host distance. Host galaxies from literature (gray circles) and this work (blue circles) are classified as either isolated (solid circles) or non-isolated (dashed circles), with circle sizes proportional to their virial radii. Satellites are displayed as colored dots within these circles, with colors showing their stellar masses and positions showing their spatial distribution on the sky. Unconfirmed satellite candidates are shown as gray crosses. We also illustrate the transition from resolved star studies ($\lesssim 4$ Mpc) to integrated light searches ($4 < D < 12\ \rm Mpc$), where our sample makes significant contributions. This visualization represents the current state of dwarf satellite surveys and highlights the diversity of satellite systems of dwarf galaxies. See Appendix \ref{sec:literature} for details on the literature compilation.
    }
    \label{fig:sats_demography}
\end{figure*}

Satellite dwarf galaxies are the smallest scale structure known in the Universe and have provided some of the most stringent tests of the Lambda Cold Dark Matter ($\Lambda$CDM) cosmological paradigm \citep{Bullock2017,Sales2022}. The number of these low-mass systems has challenged our galaxy formation models \citep[e.g., the missing satellite problem,][]{Klypin1999,Moore1999} and ultimately advanced our understanding of galaxy formation physics. Furthermore, satellite dwarf galaxies are ideal sites for investigating reionization \citep{Bullock2000}, alternative dark matter models \citep{Nadler2021}, baryonic feedback \citep{Hopkins2011}, environmental effects \citep{Wetzel2015}, and how dwarfs occupy dark matter halos \citep{Wechsler2018}.

Our best knowledge of galaxy formation and evolution in the low-mass regime comes from the satellites of our Milky Way (MW) and M31 \citep[e.g.,][]{Mateo1998,Weisz2011,McConnachie2012,Simon2019,Savino2025}. These satellites, from classical dwarfs ($M_\star \gtrsim 10^{5}\ M_\odot$) to ultra-faint dwarfs ($M_\star \lesssim 10^{5}\ M_\odot$), have been characterized in exquisite detail, showing diverse star formation histories, gas content, and chemical abundances. In recent years, satellite searches have been extended beyond the Local Group to nearby MW analogs, both through studies of individual host galaxies \citep[e.g.,][]{Danieli2017,Tanaka2018,Smercina2018,Crnojevic2019,Bennet2019,Mutlu-Pakdil2024} and systematic surveys including the Exploration of Local VolumE Satellites survey \citep[ELVES;][]{CarlstenELVES2022} and the Satellites Around Galactic Analogs survey \citep[SAGA;][]{SAGA-I,SAGA-III}. The ELVES survey studied the satellite populations of 28 MW analogs in the Local Volume ($<12$~Mpc) using surface brightness fluctuations, whereas the SAGA survey focuses on 101 MW-like systems at 25--40 Mpc using spectroscopic follow-up of satellite candidates. These efforts have allowed us to place the MW within a broader context and enabled various studies of satellite galaxies themselves \citep[e.g.,][]{Muller2018,ELVES-I,Greene2023,Danieli2023,SAGA-IV}. 

The hierarchical structure formation in $\Lambda$CDM predicts that dwarf galaxies, despite being 10-1000 times less massive than the MW, should also host their own satellite galaxies \citep[e.g.,][]{Sales2013,Wheeler2015,Dooley2017a,Dooley2017b,Munshi2019,Santos-Santos2022}. The Large and Small Magellanic Clouds (LMC, SMC) are a vivid example of the ``satellites of dwarfs" scenario in our Local Group \citep{Sales2017,Santos-Santos2021}. Exploring the satellite population of dwarf galaxies observationally will stress-test $\Lambda$CDM and galaxy formation models, and also disentangle the effects of different dark matter models from any baryonic effects that depend on the environments. 

Beyond the Local Group, significant progress has been made in identifying satellites of dwarf galaxies through deep imaging studies of individual systems \citep{Martinez-Delgado2012,Sand2015,Zhang2021_NGC6822,Kim2022,Muller2023,Sand2024,McNanna2024}. The Magellanic Analog Dwarf Companions And Stellar Halos (MADCASH) survey represents a major step in this direction, which systematically studied satellites of two LMC analogs, NGC~2403 and NGC~4214 using resolved stars with ground-based imaging data \citep{Carlin2016,Carlin2019,Hargis2019,Carlin2024}. The Large Binocular Telescope Satellites Of Nearby Galaxies Survey (LBT-SONG) further searched satellite candidates of dwarfs in both resolved stars and integrated light \citep{Garling2021,Davis2021,Davis2024}, though it lacked direct distance measurements for most candidates. More recently, the Identifying Dwarfs of MC Analog GalaxiEs (ID-MAGE) survey \citep{Hunter2025} and the LBT Imaging of Galactic Halos and Tidal Structures (LIGHTS) survey \citep{Zaritsky2024_LIGHTS} have extended these works but similarly have not provided direct distance information to confirm the physical association of satellite candidates. In Appendix \ref{sec:literature}, we present a compilation of literature results on ``satellites of dwarfs''.

Figure \ref{fig:sats_demography} summarizes the current state of this field, showing the demography of known satellite systems of dwarf galaxies. The dwarf hosts from the literature are represented by gray circles, classified as isolated (solid circles) and non-isolated (dashed circles), with circle sizes corresponding to their projected virial radii. Many searches have focused on non-isolated dwarf hosts, complicating both the association of satellite candidates with potential hosts and comparisons with theoretical predictions. Most confirmed satellites (colored dots) have been found within $\lesssim 4$~Mpc, and few low-mass hosts have been explored. Furthermore, existing surveys vary considerably in detection limits and spatial coverage. A systematic search for satellites of isolated dwarf galaxies beyond 4 Mpc is therefore urgently needed to build a statistical and uniform sample of satellite systems of dwarfs and test the $\Lambda$CDM predictions.

Satellite searches in the Local Volume typically follow a two-step procedure: detecting satellite candidates in existing imaging data, and measuring distances to confirm the association with the host. Distance information is thus critical for satellite searches since $\sim 80\%$ of the satellite candidates can be background interlopers for MW-mass hosts \citep{Carlsten2019,Bennet2019}, and it can be worse for dwarf hosts. For satellite searches in integrated light, distance measurement is challenging and typically requires space-based follow-ups to measure the tip of the red giant branch (TRGB) distance \citep[e.g.,][]{Danieli2017,Bennet2019}. Although satellite populations can be studied statistically without individual distances by subtracting background galaxy contributions \citep[e.g.,][]{Tanaka2018,Wang2021,Li2022,Bhattacharyya2024,Wang2025}, this approach brings large additional uncertainty at low mass because of higher background contamination fraction, and compromises the ability to study individual systems. Recently, the ELVES survey demonstrated that integrated light searches, combined with surface brightness fluctuation (SBF, \citealt{Tonry1988}; see \citealt{Cantiello2023} for a review) distances from ground-based imaging data, can be a very efficient way to search and confirm satellites in the Local Volume.

Encouraged by these developments, we introduce the \elvesdwarf survey, designed to create a statistical sample of satellite systems hosted by \textit{isolated} dwarf galaxies in the Local Volume ($4<D<12$~Mpc) with a well-defined selection function. We detect satellite candidates in integrated light from wide imaging surveys, then obtain SBF distances using deeper ground-based imaging data. As the first paper of the \elvesdwarf survey, we present a systematic search for satellites around 8 isolated dwarf hosts, highlighted as the blue circles in Figure \ref{fig:sats_demography}. Our sample fills a critical gap in existing observations by targeting isolated hosts at 4--12 Mpc and by including galaxies less massive than the SMC, an unexplored mass range in the literature. In this work, we utilize the archival data of the Hyper Suprime-Cam (HSC) on the 8.2 m Subaru telescope to measure SBF distances to satellite candidates. 

This paper is structured as follows. We present our dwarf host selection in Section \ref{sec:hosts} and describe satellite candidate detection in Section \ref{sec:detection}. Section \ref{sec:data} details the data used for SBF distance measurements. We then describe our SBF measurements in Section \ref{sec:sbf}. Section \ref{sec:sat_systems} describes the confirmed satellites for each host. We then present our main results in Section \ref{sec:results}, including the theoretical models, satellite abundance and mass function, and satellite quenching. We discuss the implications of these results in Section \ref{sec:discussion}. We adopt a flat $\Lambda$CDM cosmology with the present-day matter density $\Omega_m = 0.3$, dark energy density $\Omega_\Lambda = 0.7$, and a Hubble constant of $H_0 = 70\ \mathrm{km\ s^{-1}\ Mpc^{-1}}$. The photometric results are presented in the AB system \citep{Oke1983}. The stellar masses calculated in this work are based on a \citet{Kroupa2001} initial mass function. We apply MW dust extinction correction using the dust map in \citet{SFD1998} recalibrated by \citet{Schlafly2011}. Solar absolute magnitudes are taken from \citet{Willmer2018}. We adopt stellar masses of $M_\star = 2 \times 10^9\, M_\odot$ for the LMC and $M_\star = 3.2 \times 10^{8}\, M_\odot$ for the SMC \citep{Skibba2012LMC}.

\section{Dwarf Host Selection}\label{sec:hosts}
\input{hosts_cat}

Our dwarf host selection is guided by both observational constraints and scientific goals. We focus on dwarf hosts beyond 4~Mpc where satellites can be detected in integrated light, allowing us to probe a larger volume than resolved-star studies \citep[e.g.,][]{Mutlu-Pakdil2024,Carlin2024}. We target dwarf hosts with stellar masses of $7 < \log M_\star / M_\odot < 10$, complementing existing surveys of more massive hosts while ensuring that the dwarf hosts are massive enough to have a few satellites above our detection limit (see \S\ref{sec:detection}). We select only dwarf hosts with TRGB distances to ensure reliable satellite associations and physical property estimation of both the host and satellites. Different from many other studies including the MADCASH and LBT-SONG surveys, we require hosts to be \textit{isolated} to avoid ambiguity in satellite association and to make cleaner comparisons with theoretical models. As a result, we focus on the so-called ``satellite-of-dwarf'' scenario, instead of ``satellite-of-satellite.'' Our distance requirement also ensures that our hosts have not interacted with the Local Group. Most importantly, hosts must be covered by existing HSC data in the $r$ or $i$ bands for SBF distance measurements. 

Based on these criteria, we select host galaxies from the Updated Nearby Galaxy Catalog\footnote{\url{https://www.sao.ru/lv/lvgdb/}} \citep[UNGC, ][hereafter \citetalias{Karachentsev2013}]{Karachentsev2013}. UNGC compiles various galaxy properties such as distance, radial velocity, and photometry from the literature, and is updated regularly to include the latest discoveries. We use the $K_s$-band luminosity ($L_{\rm K_s}$) as a proxy for host stellar mass \citep{Bell2003} because it is relatively insensitive to extinction and young stellar populations. We thus require the hosts to have $7 < \log L_{\rm K_s} < 10$ (corresponding to $7 < \log M_\star < 10$ assuming $M_\star / L_{\rm K_s} = 1$) and have TRGB distances of $4 < D < 12$~Mpc. We also apply cuts on Galactic latitude $|b| > 15 \deg$ and extinction $E(B-V) < 0.15$ to minimize Galactic extinction.

To select isolated dwarf hosts, we make use of the ``tidal indices'', which quantify the local tidal field contributed by the neighboring galaxies \citep{Karachentsev2013}. UNGC presents three tidal indices: $\Theta_1$ only accounts for the most significant disturber; $\Theta_5$ accounts for the five most important disturbers; and $\Theta_j$ represents the average galaxy density within a 1 Mpc sphere. Negative tidal indices indicate isolation. We thus require our dwarf hosts to have negative values for all three indices. We also visually inspect each host to confirm that it is not associated with any uncatalogued groups in UNGC. This stringent isolation requirement avoids confusion about satellite ownership and indicates a relatively clean dynamical history of the host, making it easier to compare the observed results with idealized theoretical models (see \S\ref{sec:theory}). Finally, we require that the projected virial radius of each host has significant overlap with existing HSC data in either $r$ or $i$ bands (see \S\ref{sec:data} for details). We select these two filters because they have well-established SBF calibrations \citep[e.g.,][]{Carlsten2019,Carlsten2021}.

In total, there are eight dwarf hosts that satisfy all these criteria. Table \ref{tab:hosts} summarizes their key properties, including distances, heliocentric radial velocities ($v_h$), $K_s$-band luminosity ($\log L_{\rm K_s}$), $V$-band absolute luminosity ($M_V$), $g-r$ color, estimated stellar mass ($\log M_\star$), virial radius ($R_{\rm vir}$), and the unmasked fraction during detection ($f_{\rm unmask}$). The heliocentric radial velocities are from \citetalias{Karachentsev2013}, while the TRGB distances come primarily from \citet{Tully2013} and \citet{Tully2016}, except for NGC~4625 from \citet{McQuinn2017}. The unmasked fraction accounts for the fraction of areas masked by our star mask (see \S\ref{sec:detection}). Among our selected hosts, NGC~4605 was searched by \citet{Davis2024}, who reported no satellite candidates because of their small spatial coverage.

We note that UGC05427 has inconsistent TRGB distance measurements in the literature. \citet{Tully2013} reported $7.69\pm 0.10$~Mpc based on its color-magnitude diagram from the \textit{Hubble Space Telescope}, whereas \citet{Tikhonov2018} claimed 11.30~Mpc. We visually inspected the CMD of UGC05427 and found \citet{Tully2013} is more consistent with the CMD. This is further supported by an earlier distance measurement of 7.1~Mpc using the brightest blue star method \citep{Makarova1998}. We therefore adopt a distance of $7.69\pm 0.10$~Mpc for UGC05427 and have verified that the tidal indices calculated using this new distance still satisfy our isolation criteria.

Host stellar mass is important in both determining the range of satellite search and comparing observational results with theoretical predictions. The $K_s$-band luminosities listed in Table \ref{tab:hosts}, as a proxy for host stellar mass, are primarily based on shallow 2MASS data \citep{Jarrett2000,Karachentsev2013}. Several of our hosts are undetected or marginally detected in 2MASS due to their low surface brightness, but two other hosts (UGC05423 and DDO046) have much deeper NIR observations available \citep{Fingerhut2010}. Given this heterogeneity, we decide to use deep optical imaging data to obtain more consistent stellar mass estimates. The Siena Galaxy Atlas\footnote{\url{https://sga.legacysurvey.org/}} \citep[SGA,][]{Moustakas2023}, a survey presenting multi-band photometric measurements for nearby galaxies with large angular sizes, is an ideal dataset for estimating the host stellar mass. We take integrated photometry from the SGA, calculate mass-to-light ratios ($M_{\star}/L$) using the color--$M_{\star}/L_{g}$ relation from \citet{Into2013}, and convert $g$-band magnitudes to stellar masses, shown in Table \ref{tab:hosts}. Consistent stellar masses are obtained when using the \citet{Bell2003} relation. These stellar masses are lower than those derived from $L_{\rm K_s}$ assuming $M_\star / L_{\rm K_s} = 1$, but align better when using $M_\star / L_{\rm K_s} = 0.6$ or using the color--$M_{\star}/L_{\rm K_s}$ relation from \citet{Bell2003}. For the several hosts included in \citet{Leroy2019}, which present stellar masses using WISE $3.4\mu$m photometry, our estimates are higher by 0.2--0.3 dex. Considering systematic uncertainties in galaxy photometry and stellar mass estimation methods, we adopt a conservative uncertainty of 0.15 dex for the stellar masses of our dwarf hosts. This uncertainty is slightly larger than the uncertainties in \citet{Leroy2019} but is appropriate for these low-mass dwarf galaxies.

Next, we determine the radial range for our satellite search. Unlike studies of MW-like systems where satellites are searched out to a fixed radius (e.g., 300 kpc in \citealt{SAGA-III} and \citealt{CarlstenELVES2022}) to match the virial radius of the MW, our hosts span a wider stellar mass range ($10^{7.8} - 10^{9.5}\ M_\odot$), requiring a mass-dependent search radius. We calculate the halo mass using our estimated stellar mass and the stellar-to-halo mass relation (SHMR) from \citet{RP2017}, then derive the virial radius \rvir of each host. The virial radius $R_{\rm vir}$ is defined as the radius within which the average density is $\Delta=200$ times the critical density of the Universe. The resulting virial radii range from 70~kpc to 130~kpc (Table \ref{tab:hosts}). To account for uncertainties in halo mass estimates and to avoid missing satellites near the virial boundary, we search for satellite candidates within $\sim 1.1\ R_{\rm vir}$ in projection from each host. This roughly reflects the effect of stellar mass uncertainties ($\pm0.15$~dex) on the virial radius. The projected \rvir of each host is shown as black dashed circles in Figure \ref{fig:host_coverage}. 

% The Ks-band magnitude given in \citet{Karachentsev2013} is in Vega. And the Ks-band luminosity is calculated using $L_{K_s} = -0.4 * (K_s - \mathrm{DM} - K_{s, \odot})$, where $K_{s, \odot}=3.25$ is the absolute magnitude of the Sun in the $K_s$ band in Vega system. 

\section{Hyper Suprime-Cam Data}\label{sec:data}

Deep, high-resolution imaging data is essential for measuring SBF signals and associating satellite candidates with their potential hosts. We use the archival Hyper Suprime-Cam (HSC) data in the $r$ and $i$ bands, as these filters have existing SBF calibrations, superior seeing conditions, and stronger intrinsic SBF signals. 

Our HSC data come from three sources: the HSC Subaru Strategic Program \citep[HSC-SSP,][]{Aihara2022}, the HSC Legacy Archive \citep[HSCLA,][]{Tanaka2021}, and raw HSC data from the Subaru Mitaka Okayama Kiso Archive (SMOKA)\footnote{\url{https://smoka.nao.ac.jp/HSCsearch.jsp}}. While HSC-SSP covers $\sim 1200\ \mathrm{deg}^{2}$ with an impressive depth ($5\sigma$ point-source detection of $26.2\ \mathrm{mag}$ in $r$ and $i$ bands) and seeing (median seeing $\sim 0.6\arcsec$ in $i$ band), none of our hosts falls within the footprint of the 3rd public data release (PDR3) of HSC-SSP. 

HSCLA is a data archive that contains processed images from the PI-based programs outside of the HSC-SSP survey. The latest HSCLA, \code{HSCLA2016}\footnote{\url{https://hscla.mtk.nao.ac.jp/doc/}}, includes all HSC observations up to 2016 under good conditions. It covers $\sim 3400\ \mathrm{deg}^{2}$ in at least one filter with depths of 24--28~mag and seeing better than $1\farcs3$ in the $r$ and $i$ bands. The data were reduced using \code{hscPipe}, a customized version of the LSST Science Pipelines \citep{Bosch2018,Aihara2022}. Among all of the dwarf hosts, four of them have significant overlap with the HSCLA data in either $r$ or $i$ band. For the remaining four hosts, we download the raw HSC data taken after 2016 and reduce the raw HSC data taken after 2016 using the LSST Science Pipelines \citep{Juric2017,Bosch2018,Bosch2019,LSST2019,Jenness2022}\footnote{We use the \code{w\_2025\_05} version of the LSST Science Pipelines: \url{https://pipelines.lsst.io/v/w\_2025\_05/index.html}}. 

A summary of the depth, exposure time, and seeing of these data can be found in Table \ref{tab:obs}. Although previous HSC-based SBF studies primarily used $i$-band data, most of our hosts have only $r$-band data available. The depth of our HSC data is comparable to or exceeds that of HSC-SSP data previously used for SBF distance measurements of Local Volume dwarfs \citep[e.g.,][]{Kim2021,CarlstenELVES2022,Kim2022}.

HSC data products include two types of coadd images, one with local sky background subtraction and one with ``global'' sky subtraction. The local sky subtraction produces over-subtraction around bright sources, whereas the global sky subtraction better preserves the low surface brightness outskirts of extended objects (see \citealt{Li2021,Aihara2022}). We note that an over-subtracted image would bias the total flux low and, in turn, bias the SBF signal high (see \S\ref{sec:sbf} for details). Therefore, in this work, we use the coadd images with global sky subtraction to measure the SBF distance. We also use the coadd PSF model at the location of the galaxy when measuring SBF.

\begin{deluxetable}{lccclc}
\tablecaption{Summary of HSC Data\label{tab:obs}}
% \tablewidth{0pt}
\tablehead{
    \colhead{Host} & 
    \colhead{Data} &
    \colhead{Band} &
    \colhead{Depth} & 
    \colhead{$t_{\rm exp}$} &
    \colhead{Seeing}
}
\startdata
UGC05423 & HSCLA & $i$ & 27.0 & 800 s & $0.65\arcsec$ \\
UGC00685 & HSCLA & $r$ & 26.5 & 800 s & $0.60\arcsec$ \\
NGC5238 & HSCLA & $r$ & 26.3 & 390 s & $0.45\arcsec$ \\
NGC4707 & HSCLA & $r$ & 26.5 & 390 s & $0.52\arcsec$ \\
NGC4605 & HSC & $r$ & 26.2 & 1530 s & $1.15\arcsec$ \\
DDO046 & HSC & $r$ & 26.4 & 710 s & $0.75\arcsec$ \\
NGC4625 & HSC & $i$ & 27.1 & 5250 s & $0.87\arcsec$ \\
UGC05427 & HSC & $r$ & 26.9 & 1010 s & $0.69\arcsec$ \\
\enddata
\tablecomments{Summary of the HSC data used in this work for each host, including the data source (HSCLA or custom-reduced HSC), filter, depth, exposure time ($t_\mathrm{exp}$), and median seeing. The depth is measured as the 5$\sigma$ point-source detection limit in AB magnitudes, similar to \citet{Aihara2022}. The HSC data used in this work is comparable to or deeper than the HSC-SSP data.}
\vspace{-2em}
\end{deluxetable}

\section{Satellite Candidate Detection and Photometry}\label{sec:detection}

\begin{figure*}
    \centering
    \includegraphics[width=1\linewidth]{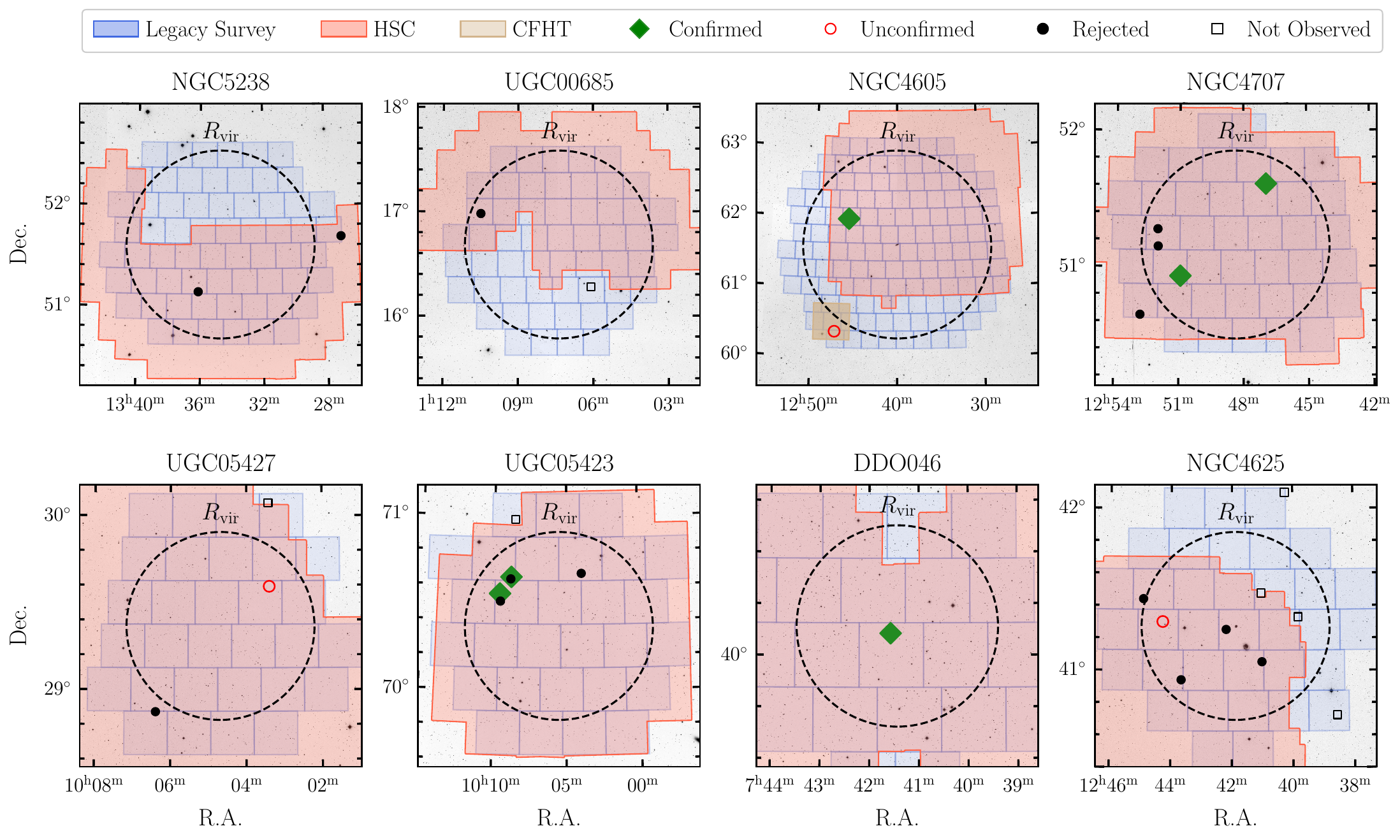}
    \caption{Survey footprint of the eight dwarf hosts in our sample. For each host, the coverage of the Legacy Survey (used for satellite candidate detection) is shown in blue, and HSC coverage (used for SBF) is shown in red, with the projected virial radius \rvir highlighted by a black dashed circle. We also mark the positions of the confirmed satellites (green diamonds), unconfirmed satellite candidates (red open circles), rejected candidates (black-filled circles), and candidates without HSC data (open squares).}
    \label{fig:host_coverage}
\end{figure*}
\subsection{Detection}

For each dwarf host in Table \ref{tab:hosts}, we detect satellite candidates using the Legacy Surveys Data Release 10 data\footnote{\url{https://www.legacysurvey.org/dr10/description/}} \citep{Dey2019}. The Legacy Surveys include the Beijing-Arizona Sky Survey \citep[BASS,][]{BASS}, covering the northern sky at $\mathrm{Dec} > +32^{\circ}$, the Dark Energy Camera Legacy Survey (DECaLS) covering the southern sky, and the additional public data taken with the Dark Energy Camera \citep{DECam}. BASS has shallower depth and worse seeing than the other components of the Legacy Surveys. Most of our hosts fall within the BASS footprint, with only UGC00685 and UGC05427 covered by DECaLS. We first download the coadd images in the $g$ and $r$ bands for the ``bricks'' that overlap with the searching radius of each host. We then apply our detection pipeline, which is based on the low-surface-brightness galaxy detection algorithm developed by \citet{Greco2018} and further improved by \citet{Carlsten2021} and \citet{CarlstenELVES2022}. Here, we briefly summarize the key steps of this pipeline and refer interested readers to \citet{Greco2018}, \citet{Li2022}, and \citet{CarlstenELVES2022} for details. 

Our pipeline is optimized for detecting extended low surface brightness objects while also preserving higher surface brightness candidates. The process begins by masking stars brighter than 19~mag in both $g$ and $r$ bands by cross-matching with Gaia DR2 catalogs \citep{GAIA2018}. As listed in Table \ref{tab:hosts}, about 94\% of the area remains after masking the bright stars. Then, we identify and replace bright sources and their associated diffuse light with sky noise by applying surface brightness thresholding. Additionally, we use \code{sep} \citep{Barbary2016} with a small mesh size to remove compact sources and noise peaks. A ``cleaned'' image is produced up to this step where only extended low surface brightness features remain. To further emphasize these features, we smooth the cleaned images with a Gaussian kernel with FWHM $\sim 2.5-3.5$ times the PSF FWHM. We then run \code{SExtractor} \citep{Bertin1996} on both the smoothed and cleaned images in both bands using a Gaussian kernel with FWHM$=5 \,\mathrm{pix}$ and a low detection threshold of $\sim 2-2.5\sigma$. We require the final candidates to be detected in both bands, to remove artifacts such as chip edges or bright star ghosts. The specific configuration used for each host varies slightly to accommodate different data quality and is determined by trial and error to maximize the detection completeness. We note that our detection limit is slightly deeper than that of the ELVES survey to ensure that we capture the very few satellites of dwarf hosts. This inevitably increases the number of false positives. 

To eliminate false positives, we visually inspect all detections, removing artifacts (e.g., shredded stellar wings, galaxy outskirts, and coadding issues) and obvious background galaxies (e.g., merging systems, tidal features, edge-on galaxies, and those with color gradients or spiral arms; see the Appendix in \citealt{Carlsten2020}). We make use of existing velocity measurements from NED \citep{NED}, SIMBAD \citep{SIMBAD} and DESI DR1 \citep{DESI-DR1} and remove objects with radial velocities offset by $\Delta v_h > 300\ \mathrm{km\ s^{-1}}$ from the host. Less than 5\% of initial detections survive this visual inspection. Among the remaining objects, many are still background galaxies resembling dwarfs in the Legacy Surveys data, because of the limited depth and resolution. However, these objects will manifest themselves as outliers of the average mass--size relation of satellites assuming that the candidates are all at the distance of the host. We thus place the candidates on the mass--size plane using the photometry from \S\ref{sec:photometry} and reject objects that are $>2\sigma$ away from the average mass--size relation of satellites from \citet{ELVES-I}\footnote{95\% of the real satellites should be included after this cut, assuming that the mass-size relation of satellites of dwarfs is the same as satellites of MW analogs. This is a valid assumption given the lack of evidence for environmental dependence in the mass-size relation \citep{ELVES-I}.}. This step effectively removes objects that are too compact for their luminosity at the host distance. Roughly 50\%--70\% of the candidates are removed in this step. We then perform a final round of visual inspection using HSC data for objects with HSC coverage (see \S\ref{sec:data}). In the end, each host has 1--9 satellite candidates. Figure \ref{fig:host_coverage} shows the spatial distribution of these candidates, where some of the candidates are outside the projected \rvir. In \S\ref{sec:sbf}, we measure the SBF distances for these satellite candidates using deep, high-resolution HSC data (\S\ref{sec:data}).

% Most of the objects appear smooth and featureless in the Legacy Surveys data. 

Understanding our detection completeness is essential to determine the mass limits of the satellites in our survey and to make meaningful comparisons with theoretical models. Following \citet{CarlstenELVES2022} and \citet{Li2022}, we quantify completeness by injecting mock dwarf galaxies into Legacy Surveys data and recovering them with our detection algorithm. More details can be found in Appendix \ref{sec:completeness}. The detection completeness of each host as a function of the total magnitude and surface brightness is shown in Figure \ref{fig:completeness}. For hosts at $D<8$~Mpc, we achieve $>50\%$ completeness to $M_g \approx -8.5$ ($M_\star \approx 10^{5.4}\, M_\odot$ assuming an average color of $g-r=0.4$); while for the three more distant hosts ($D>8$~Mpc), our $50\%$ completeness limit is $M_g \approx -9.5$ ($M_\star \approx 10^{5.7}\, M_\odot$). This completeness is comparable to that of the ID-MAGE survey \citep{Hunter2025}, and is slightly higher than that reported in \citet{CarlstenELVES2022} for hosts at similar distances due to the use of a larger smoothing scale and lower detection threshold. The HSC data we use for measuring SBF are deep enough, such that we expect no additional incompleteness from the quality of the HSC data.

\subsection{Photometry}\label{sec:photometry}
We measure the structural and photometric properties of the satellite candidates by fitting a single \sersic model to the Legacy Surveys $g$ and $r$-band coadd images, following a similar approach to \citet{ELVES-I}, \citet{CarlstenELVES2022}, and \citet{Li2024}. Most satellite candidates appear smooth without showing SBF signals in the Legacy Surveys data and are well-fit by a single \sersic model. Using \code{IMFIT} \citep{imfit}, we first fit a \sersic model to the $g$-band data, allowing the central R.A. and Dec., total magnitude ($m_g$), effective radius ($r_e$), \sersic index ($n$), and ellipticity ($\varepsilon$) to vary. The effective radius $r_e$ is defined as the half-light radius along the semi-major axis. Then, we fit the $r$-band data using the best-fit model from the $g$ band while only allowing the total magnitude ($m_r$) to vary.
% We use \code{sep} to generate segmentation maps, which are used to mask out other sources during \sersic fitting. 
Table \ref{tab:sats} presents the photometric properties of all satellite candidates. We only use the Legacy Surveys for photometry rather than the deeper HSC data because not all candidates have HSC coverage in consistent filters. Photometric uncertainties are estimated following \citet{ELVES-I} by injecting and recovering mock galaxies into the Legacy Surveys data (see Appendix E in \citealt{ELVES-I}). We derive stellar masses for these candidates assuming that they share the same distance as their hosts, and using the color--$M_\star/L$ relation from \citet{Into2013}. We compare our photometry with results from the SMUDGes survey \citep{Zaritsky2023} and find good agreement. For a few bright satellite candidates that cannot be described by a single \sersic model, we take the photometric properties from the SGA survey \citep{Moustakas2023}. 

Integrated color (such as $g-r$) is required for SBF measurements. For satellite candidates well-described by a \sersic model, we use the $g-r$ color from the \sersic fitting for SBF measurement. However, for several bright dwarfs where a single \sersic model is not capable of describing their light distribution (e.g., UGC05455, a satellite candidate of UGC05423), we measure the surface brightness profiles by doing isophote fitting using \code{Photutils}\footnote{\url{https://photutils.readthedocs.io/en/stable/reference/isophote\_api.html}}, similar to the approach used in SGA \citep{Moustakas2023}. We estimate the $g-r$ color within the region where we measure SBF signals and use that color for SBF measurement.

\section{SBF Distances to Satellite Candidates}\label{sec:sbf}

\begin{figure*}
    \centering
    \includegraphics[width=1\linewidth]{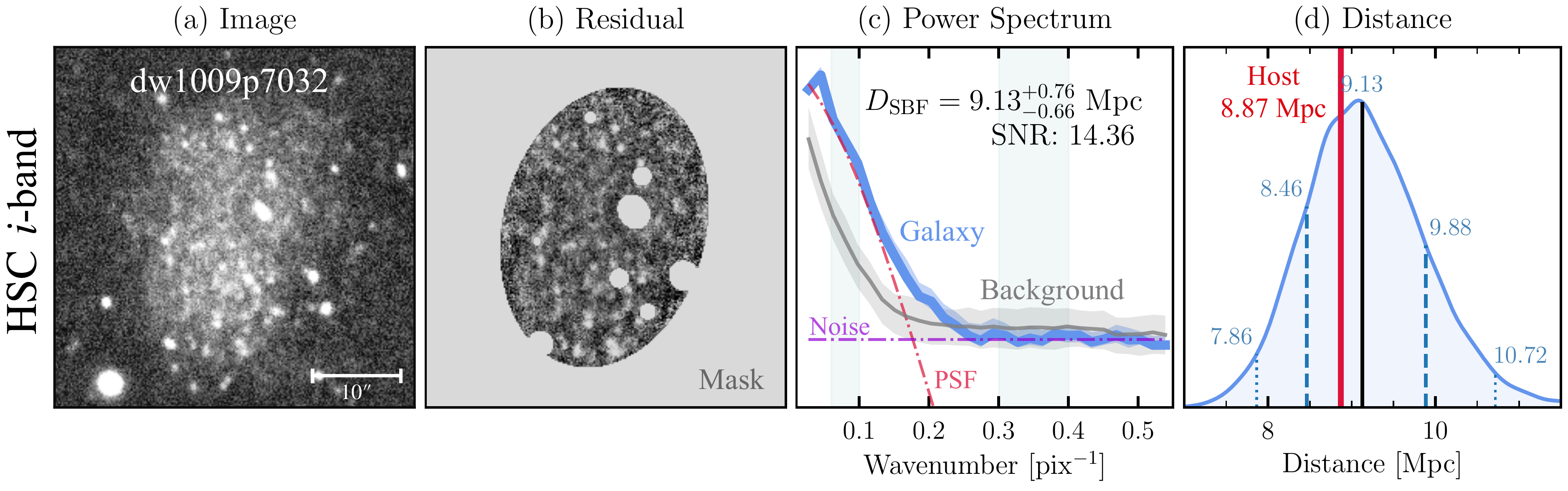}
    \caption{An example SBF distance measurement. We show the SBF measurement process for dw1009p7032, a confirmed satellite galaxy associated with UGC05423 ($D=8.87$~Mpc). Panel (a) shows the coadded HSC $i$-band image of dw1009p7032. Due to its non-\sersic morphology, we construct a smooth galaxy model using median filtering with a window size of 10 times the seeing. We then subtract the smooth model from the image and divide the residual by the square root of the smooth model. We also mask out contaminating compact sources brighter than $M_{i,\rm thresh}=-4.8$~mag. Panel (b) shows this masked residual image. In Panel (c), we calculate the azimuthally averaged power spectrum of the residual image (blue line) and fit it with a combination of the PSF power spectrum (red line) and a constant white noise (purple line). To account for the contribution from unmasked sources, we measure the SBF signal in the background fields (gray-shaded region) and subtract it from the SBF signal from the galaxy. Using the SBF--color relation, the resulting SBF distance is $D_{\rm SBF} = 9.13^{+0.76}_{-0.66}\ \rm Mpc$ with $\mathrm{S/N}=14.36$. Panel (d) shows the derived SBF distance distribution, with the median, $1\sigma$, and $2\sigma$ confidence levels marked. The distance of the host galaxy is highlighted in red. The distance of dw1009p7032 is fully consistent with it being a satellite of UGC05423.
    }
    \label{fig:sbf_demo}
\end{figure*}

Distance measurement is essential for confirming associations between satellite candidates and their potential dwarf hosts. In this section, we measure the SBF distances of satellite candidates using HSC data and associate them with their hosts. Our methodology closely follows the approaches of \citet{Carlsten2019}, \citet{Carlsten2021}, \citet{CarlstenELVES2022}, and \citet{Li2024}. We suggest readers interested in detailed SBF measurement techniques read these references.

\subsection{SBF Measurement}\label{sec:sbf_meas}

SBF quantifies the pixel-to-pixel fluctuations in the number of bright stars within a galaxy. As distance increases, each resolution element includes more stars, and the variance decreases. Thus, SBF is sensitive to both distance and the stellar population of the galaxy. Furthermore, the fluctuation caused by background galaxies and other unresolved sources (such as globular clusters) affects the measured fluctuation, requiring masking and additional statistical corrections. Optical colors, as a proxy for the underlying stellar population, are typically used to calibrate the absolute SBF magnitudes. In this work, we use the SBF calibration in \citet{Carlsten2019} for the $i$ band. For the $r$ band, we have recalibrated the SBF--color relation using data from \citet{Carlsten2021}, as detailed in Appendix \ref{sec:sbf_calib_r}.

An example SBF measurement is shown in Figure \ref{fig:sbf_demo}. We start measuring the SBF distance by constructing a ``residual image'', defined as $\rm (Galaxy - Model) / \sqrt{Model} \times Mask$, where `Galaxy' is the observed galaxy image, `Model' is a smooth model describing the light distribution of the galaxy, and `Mask' is a binary mask accounting for objects that contribute to the fluctuation but are not intrinsic to the target galaxy (e.g., globular clusters and background galaxies). This residual image contains all the necessary information about the SBF signal.

We employ two different methods to construct a smooth galaxy model depending on the specific case. For galaxies with regular morphology, we fit a single \sersic model using \code{IMFIT} \citep{imfit}. For galaxies with non-\sersic morphology, we adopt the method in \citet{Kim2021} and use a median filter with a size of 5--10 times the seeing to obtain a smooth model. We adjust the median filter size on a case-by-case basis, aiming to preserve high-frequency SBF signals while removing the overall low-frequency structures. As demonstrated in \citet{Kim2021}, these two methods agree well except for the largest spatial scales ($k<0.03\ \mathrm{pix}^{-1}$). For highly elongated objects, we only use \sersic fitting since the median-filter method performs poorly in such cases. As noted in \S\ref{sec:data}, if the image suffers from sky over-subtraction issues, the smooth model will be underestimated, which in turn increases the amplitude of the residual image and biases the measured distance toward lower values. Therefore, we recommend using global sky subtraction when reducing the image to ensure accurate SBF measurements, especially for bright and extended sources. 

To mask globular clusters and background galaxies, we use an absolute magnitude threshold $M_{\rm thresh}$, above which sources are masked \citep{Carlsten2019}, assuming the target galaxy is at the distance of the potential host. This threshold is motivated by the need to mask out sources brighter than the brightest RGB star in a galaxy (which is about $M_i \approx -4.5$~mag and $M_r\approx -4$~mag) while keeping the RGB stars intact. In practice, the masking threshold is determined for each target galaxy to ensure that bright, compact, non-SBF features are properly masked. The threshold is in a range of $-5 \lesssim M_{\rm thresh} < -4$ for both $r$ and $i$ bands, with the $r$-band threshold generally being fainter than the $i$-band threshold because RGB stars are fainter in the $r$ band. To enhance the SBF signal-to-noise ratio (S/N), we mask the area where the surface brightness falls below $f_{\rm mask}$ times the central surface brightness, with $f_{\rm mask}$ randomly drawn from a uniform distribution $f_{\rm mask}\sim \mathcal{U}(0.15,0.25)$ to determine the SBF uncertainty. For galaxies with central star-forming regions or nuclear star clusters, we also mask the central regions. For bright galaxies, we measure the SBF signal in the outskirts where it is not dominated by star-forming regions. With both the smooth model and the mask in hand, we construct the residual image. An example is shown in the second panel of Figure \ref{fig:sbf_demo}, where the masked region is highlighted in gray. 

To extract the SBF signal, we compute the azimuthally averaged power spectrum of the residual image. This power spectrum $P(k)$ can be expressed as a linear combination of the PSF power spectrum and a white noise floor: 
\begin{equation}\label{eq:sbf}
    P(k) = f_{\rm SBF} \cdot E(k) + P_{\rm noise},
\end{equation}
where $f_{\rm SBF}$ is the SBF signal that we aim to measure, and $E(k)$ is the power spectrum of the PSF convolved with the mask's power spectrum. We refer the readers to \citet{Liu2000} for a more rigorous derivation of this relation. The distance modulus can then be determined as $\mu_{\mathrm{DM}} = \overline{m}_{\rm SBF} - \overline{M}_{\rm SBF}$, where $\overline{m}_{\rm SBF} = -2.5\log_{10}(f_{\rm SBF}) + \mathrm{ZP}$ is the observed SBF magnitude, and $\overline{M}_{\rm SBF}$ is the absolute SBF magnitude derived from the SBF--color relation. The photometric zeropoint is $\mathrm{ZP} = 27.0$~mag for HSC data. The azimuthally averaged power spectrum of the residual image is shown as the blue line in the third panel of Figure \ref{fig:sbf_demo}. We then fit it with a combination of the PSF's power spectrum (red dash-dotted line) and a constant white noise (purple dash-dotted line). The wavenumber range for fitting is $k_1 < k < k_2$ where $k_1\sim \mathcal{U}(0.05, 0.10)\ \mathrm{pixel}^{-1}$ and $k_2 \sim \mathcal{U}(0.3, 0.4)\ \mathrm{pixel}^{-1}$. The light green vertical bands in Figure \ref{fig:sbf_demo} represent the possible ranges for $k_1$ and $k_2$. 

To correct for the contribution from the unmasked background sources, we select several relatively empty fields around the target galaxy and construct the residual image as $\rm Background / \sqrt{Model} \times Mask$, where `Model' is the same as that of the target galaxy, and `Mask' is generated using the same $M_{\rm thresh}$. The SBF signal from the background fields is measured similarly, shown as the gray shaded region in Figure \ref{fig:sbf_demo}. The SBF signal from the background is then subtracted from the galaxy SBF signal. 

The uncertainty of the measured SBF signal is derived by doing 100 Monte Carlo random draws for $k_1$, $k_2$, $f_{\rm mask}$, and background fields \citep{Cohen2018}. We also define the SBF S/N as the median SBF signal divided by its uncertainty. The example in Figure \ref{fig:sbf_demo} has a high S/N of 14. In the end, we convert $\overline{m}_{\rm SBF}$ and its uncertainty to distance using the SBF--color relations, taking into account the color uncertainty and the scatter of the SBF calibration (see Appendix \ref{sec:sbf_calib_r}). For bright galaxies exhibiting color gradients from the center to the outskirts, we use the average color in the region where we measure the SBF. Panel (d) of Figure \ref{fig:sbf_demo} shows the measured SBF distance distribution with median, $1\sigma$ (68\%), and $2\sigma$ (95\%) ranges marked with blue vertical lines, and the host distance marked in red.

One of our hosts, UGC05423, is partially covered by galactic cirrus. Measuring SBF distances directly from the coadd image would absorb the flux from the cirrus into the smooth galaxy model, resulting in overestimated distances. To address this, we attempt to remove the cirrus by aggressively masking out sources (including the target galaxy) and then median-filtering the masked image with a kernel size of 5 times the seeing. We validate this approach by injecting mock galaxies into cirrus-heavy fields using \code{ArtPop} \citep{artpop} and recovering the underlying SBF signals. Our tests indicate that this method effectively removes the cirrus contamination to a satisfactory degree. Our tests show that while the cirrus subtraction introduces additional uncertainties, they remain minor compared to other sources of error, such as the galaxy color and masking.

\subsection{Group Association}\label{sec:psat}
Following \citet{Carlsten2021}, we classify the satellite candidates into three categories based on their SBF distance measurements. We consider a candidate as a ``confirmed'' satellite if it has an SBF measurement with a high signal-to-noise ratio ($\mathrm{S/N}>5$) and its SBF distance is consistent with the host within $2\sigma$. If the lower limit of the $2\sigma$ range is beyond the host distance, we reject the candidate and consider it a background galaxy. If a candidate has a low S/N ($<5$) but its lower $2\sigma$ limit is below the host distance, we classify it as ``unconfirmed''. Deeper and higher-resolution data are required to determine the status of an unconfirmed candidate. Candidates outside the HSC footprint are also considered as ``unconfirmed''. The $\mathrm{S/N}>5$ cut ensures that the fluctuation comes from SBF rather than other sources, such as star-forming regions. This threshold has been validated against HST TRGB distances \citep{Bennet2019} and has been applied in the ELVES survey \citep{Carlsten2021,CarlstenELVES2022}.

In addition to SBF, we also utilize radial velocity data when available to confirm satellite associations. We search for velocity measurements primarily in the SIMBAD database and DESI DR1 \citep{DESI-DR1}. A satellite candidate is considered confirmed if its radial velocity is within $200\ \mathrm{km\ s^{-1}}$ of its potential host galaxy, and is rejected otherwise. We note that the velocities are mostly helpful in identifying background galaxies rather than finding real satellites. 

For the unconfirmed candidates (including those without HSC data), we assign a satellite probability ($P_{\rm sat}$) based on the absolute magnitude and surface brightness, following \citet{CarlstenELVES2022}. They showed that the confirmed satellites occupy distinct regions in the absolute magnitude--surface brightness space, whereas rejected candidates are typically smaller and with higher surface brightness. This $P_{\rm sat}$ is empirically estimated based on the confirmed and rejected candidates in the ELVES survey (see \S 5.6 in \citealt{CarlstenELVES2022}). We also assign $P_{\rm sat}=1$ for confirmed satellites and $P_{\rm sat}=0$ for rejected satellites. We note that the $P_{\rm sat}$ model for satellites of dwarf hosts might be different from that of MW analogs. We will revisit this when we have a larger sample of both confirmed and rejected objects.

\section{Satellite Systems}\label{sec:sat_systems}

\begin{figure*}[htbp!]
    \centering
    \includegraphics[width=1\linewidth]{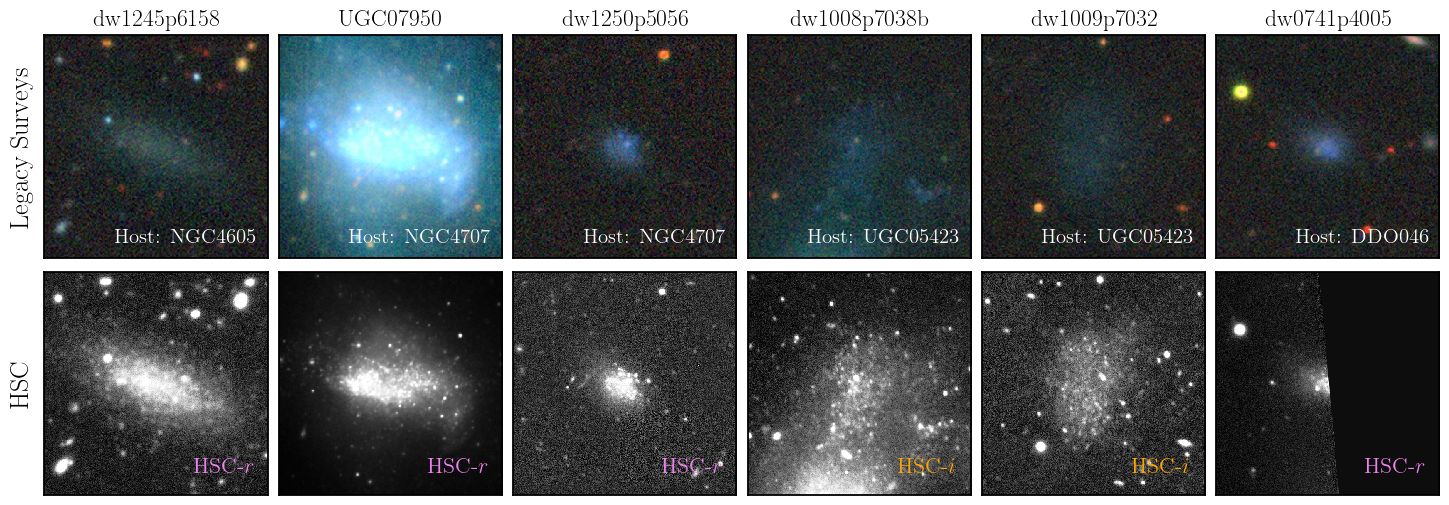}
    \caption{Cutout images of all confirmed satellites. The top panels show the color-composite images from the Legacy Surveys DR10, and the bottom panels show the single-band data from HSC. These images have 1\arcmin{} on a side and are stretched to show the SBF signals.}
    \label{fig:stamps}
\end{figure*}

In this section, we present the distance measurements for satellite candidates of each host. Table \ref{tab:sat_status} summarizes our satellite systems. Of all satellite candidates around eight dwarf hosts, we confirm six as real satellites based on our distance measurements. We reject 14 candidates as background objects, while 10 remain unconfirmed due to either shallow data or lack of HSC observations. A comprehensive catalog of all satellite candidates, including their measured properties, appears in Table \ref{tab:sats}. 

We present the cutout images of all confirmed satellites in Figure \ref{fig:stamps}. The top panels show the color-composite images from the Legacy Surveys DR10, and the bottom panels show the single-band images from HSC. We discuss each host system below, arranged by increasing host stellar mass. 

\setlength{\tabcolsep}{3pt}
\begin{deluxetable}{lccccc}
\tablecaption{Overview of Satellite Systems \label{tab:sat_status}}
\tablewidth{0pt}
\tablehead{
    \colhead{Host} & 
    \colhead{Cand.} &
    \colhead{Conf.} &
    \colhead{Reject.} & 
    \colhead{Unconf.} &
    \colhead{$N_{\rm sat,\, max}$}
}
\startdata
NGC5238 & 2 & 0 & 2 & 0 & 0 \\
UGC00685 & 2 & 0 & 1 & 1 & 0.05\\
NGC4605 & 2 & 1 & 0 & 1 & 1\\
NGC4707 & 5 & 2 & 3 & 0 & 2\\
UGC05427 & 3 & 0 & 1 & 2 & 0.1\\
UGC05423 & 6 & 2 & 3 & 1 & 2\\
DDO046 & 1 & 1 & 0 & 0 & 1\\
NGC4625 & 9 & 0 & 4 & 5 & 1.3\\
\enddata
\tablecomments{Summary of the satellite systems of dwarf hosts in this work. For each host, we list the number of satellite candidates, confirmed satellites, rejected satellites, and unconfirmed candidates (including those with no HSC observations). The last column shows the upper limit of satellite abundance $N_{\rm sat}$, calculated by summing up the satellite probability $P_{\rm sat}$ of all candidates within $R_{\rm vir}$ (see \S\ref{sec:sbf}). A complete list of satellite candidates can be found in Table \ref{tab:sats}.}
\vspace{-3em}
\end{deluxetable}

\subsection{DDO046}
DDO046, also known as UGC03966, is located at $D_{\rm TRGB}=10.38\pm0.10\ \rm Mpc$ \citep{Tully2013} and has a stellar mass about 1/5 that of the SMC. We identified only one satellite candidate, dw0741p4005, located very close to the host. It is partially covered by HSC $r$-band data and shows strong SBF signals. We measured its SBF distance to be $D_{\rm SBF}=10.28^{+2.57}_{-1.96}\ \rm Mpc$, being consistent with the host distance. The large SBF distance uncertainty comes from the fact that it has a very blue color ($g-r=0.165$), where the SBF--color relation has a large scatter at the blue end.

\subsection{UGC05427}
UGC05427 is located at $D_{\rm TRGB}=7.69\pm 0.10\ \rm Mpc$ \citep{Tully2013} with a stellar mass 1/4 that of the SMC. We found three satellite candidates, but only dw1003p2935 is within the projected $R_{\rm vir}$. dw1003p2935 shows SBF signals, but its lower limit of $2\sigma$ confidence level is higher than the host distance. We suspect that the fluctuation comes from the star-forming regions. We thus classify this candidate as ``unconfirmed.'' Future observations are needed to determine its status. The second candidate, dw1006p2852, is very smooth in the HSC data and is rejected. The third candidate, dw1003p3004, is not covered by HSC data. It falls onto chip edges in the Legacy Surveys data with low image quality but shows some hints of a bar and a disk component. We thus assign a satellite probability based on its absolute magnitude and surface brightness (see \S\ref{sec:psat}).

\subsection{NGC~5238}
NGC~5238 is located at $D_{\rm TRGB}=4.51\pm 0.10$~Mpc \citep{Tully2013}, with a stellar mass 1/4 that of the SMC. We found two satellite candidates, one within the projected $R_{\rm vir}$, and the other one outside $R_{\rm vir}$. We measured SBF distances to these two candidates using HSCLA $r$-band data and rejected both candidates. They have a smooth morphology in HSCLA data. Thus, we find no confirmed satellites associated with NGC~5238.

\subsection{UGC00685}
UGC00685 is located at $D_{\rm TRGB}=4.71 \pm 0.08$~Mpc \citep{Tully2013}. This field is contaminated by foreground galactic cirrus. We identified two satellite candidates, but only one of them (dw0110p1658) is covered by HSCLA data. It shows no SBF signal and is rejected as a real satellite. For the one without HSCLA data, it has a $P_{\rm sat}=0.05$, very unlikely to be a real satellite.

\subsection{NGC~4605}
NGC~4605 is a dwarf galaxy at $D_{\rm TRGB}=5.55\pm0.10$~Mpc \citep{Tully2013} with a similar stellar mass as the LMC. This host is also included in the ID-MAGE survey \citep{Hunter2025}, which discovered four high-likelihood candidates, but their information is not publicly available yet. 

We identified two satellite candidates. The first candidate dw1245p6158 is a dwarf spheroidal galaxy that shows a booming SBF signal. It was also discovered by \citet{Karachentsev2025a} (Dw1245+61), who obtained a relatively low S/N optical spectrum and measured the radial velocity to be $v_h = 68\pm 40\ \rm km\ s^{-1}$ based on absorption lines, being $\Delta v_h \approx 110\ \rm km\ s^{-1}$ from the host. We consider this velocity to be unreliable due to the low S/N spectrum. We measured the SBF distance to this satellite candidate using the HSC $r$-band data and got a distance of $D_{\rm SBF} = 4.72^{+0.56}_{-0.61}\ \rm Mpc$. This distance measurement confirms dw1245p6158 as a bona fide satellite of NGC~4605.

The second candidate, dw1247p6021, is not covered by HSC data. We thus use the $r$-band image from the archival Canada-France-Hawaii Telescope (CFHT)/MegaCam data\footnote{\url{https://www.cadc-ccda.hia-iha.nrc-cnrc.gc.ca/}}, following the same methodology in \citet{CarlstenELVES2022}. Because these data are quite shallow compared with the HSC data, we do not detect a significant SBF signal and are unable to either confirm or reject this candidate. We thus classify this candidate as unconfirmed.

\subsection{NGC~4707}
NGC~4707 is a dwarf spiral galaxy located at $D_{\rm TRGB}=6.52\pm0.10\ \rm Mpc$ \citep{Tully2013} and having a radial velocity of $v_h=468\ \mathrm{km\ s^{-1}}$. It is slightly less than the SMC. We found 5 satellite candidates around NGC~4707, with one candidate (dw1252p5038) being outside $R_{\rm vir}$. Among the remaining candidates, dw1250p5056 shows prominent SBF signals and has an SBF distance of $D_{\rm SBF} = 7.81^{+1.81}_{-1.31}\ \rm Mpc$, being confirmed as a satellite in this group. 

Notably, a bright star-forming galaxy, UGC07950, is $\sim30$ arcmin away from NGC~4707 and has a radial velocity of $v_h=508\ \mathrm{km\ s^{-1}}$, consistent with being a group member. To further confirm its association with NGC~4707, we measured the SBF distance using the outskirts of UGC07950. We carefully masked the central star-forming region and used the median-filtering method to construct the smooth galaxy model. Using isophotal analysis, we measured the color profile and derived an average color in the unmasked region ($g-r=0.34\pm0.07$~mag). The resulting SBF distance is $D_{\rm SBF} = 7.04^{+1.21}_{-0.84}\ \rm Mpc$, fully consistent with the distance of NGC~4707. 

Because of the complex morphology of UGC07950, we do not fit a \sersic model to estimate its photometric property. Instead, we adopt the measurements from the isophotal analysis in SGA \citep{Moustakas2023}. Assuming UGC07950 is at the same distance as NGC~4707, it has a stellar mass of $\log M_\star/M_\odot \approx 7.94$, only slightly lower than the stellar mass of the host. Thus, we consider this group as a dwarf galaxy pair. In total, two satellites (dw1250p5056 and UGC07950) are confirmed in this group. 

\subsection{UGC05423}
UGC05423, also known as M81-dwB, is at $D_{\rm TRGB}=8.87\pm 0.10\ \rm Mpc$ with a similar stellar mass as the SMC. It is not a satellite of M81 ($D\approx 3.7$~Mpc), despite as indicated by its name, but rather a galaxy in the background \cite{Chiboucas2013,Okamoto2015,Bell2022}. The field surrounding this host is contaminated by galactic cirrus as discussed in \S\ref{sec:sbf_meas}. 

We identified six satellite candidates around UGC05423. We note that d1006+69, which is an ultra-faint satellite candidate of the M81 group \citep{Okamoto2015,Gozman2024}, is in our search region but is not detected due to its low surface brightness and luminosity. Among the satellite candidates, three are rejected based on their SBF distances. One candidate (dw1008p7058) lies outside of $R_{\rm vir}$ and is not covered by HSC. This candidate is a low surface brightness blob and shares a similar color as the nearby galactic cirrus. Future observations are needed to unveil its nature. 

Another candidate, dw1009p7032, shows strong SBF signals in the HSCLA $i$-band data (see Figure \ref{fig:sbf_demo}). We measured its SBF distance to be $D_{\rm SBF} = 9.13^{+0.76}_{-0.66}\ \rm Mpc$, confirming it as a real satellite. This satellite was also reported by \citet{Chiboucas2013}, who obtained HST ACS images and measured its CMD. They did not detect a red giant branch in the CMD and concluded that it is not associated with the M81 group. Nevertheless, \citet{Karachentsev2013} claimed that dw1009p7032 is at 9.0~Mpc based on the same CMD data, in agreement with our SBF distance. 

The last candidate, dw1008p7038 (UGC05455), is a bright star-forming dwarf galaxy sitting 23 arcmin away from UGC05423. It has a radial velocity of $v_h = 1288\ \kms$ \citep{Schneider1992}, differing from the host by $940\ \mathrm{kms^{-1}}$. Despite showing prominent SBF signals in HSCLA data, our SBF distance measurement shows $D_{\rm SBF} = 14.62^{+1.54}_{-1.42}\ \rm Mpc$, leading us to classify UGC05455 as a background galaxy. 

Intriguingly, we found another object to the northwest of UGC05455, showing stronger SBF signals than UGC05455 itself. This new galaxy, dw1008p7038b, is superposed on the outskirts of UGC05455 and is bluer than UGC05455. We then consider this galaxy as a separate satellite candidate and measure its SBF distance to be $D_{\rm SBF} = 10.63^{+1.49}_{-1.16}\ \rm Mpc$, fully consistent with being a real satellite. A more detailed description of this system, together with the HSCLA image, can be found in Appendix \ref{ap:UGC05455}. To summarize, two satellites (dw1009p7032, dw1008p7038b) are confirmed to be associated with this host.

\subsection{NGC~4625}
NGC~4625 is a spiral dwarf at a distance of $D_{\rm TRGB}=11.75\pm 0.50\ \rm Mpc$ \citep{McQuinn2017} with a similar mass as the LMC. We found 9 satellite candidates. Only five of them are covered by the HSC $i$-band data. Four of them are rejected as background galaxies, but dw1242p4115 (also known as N4625A in \citealt{Karachentsev2015_NGC1156}) is found to be a foreground dwarf galaxy at $D_{\rm SBF} = 7.96^{+0.76}_{-0.77}\ \rm Mpc$. It is most likely to be a satellite of NGC~4618, an LMC analog at a distance of $D=7.66\ \rm Mpc$ \citep{Karachentsev2018}. We will do a comprehensive search around this host in the future. For satellites lacking HSC data, we assign $P_{\rm sat}$ to them. 

\section{Results}\label{sec:results}
In this section, we calculate the satellite abundance and the satellite stellar mass function for each system and compare them with the literature results and the predictions from theoretical models. We also show the color distribution of the confirmed satellites and discuss the satellite quenched fraction. We note that the area lost to bright star masks is quite small (6\% on average) and is thus not corrected for when calculating the satellite abundance and mass function. We do not correct for detection completeness either since most of the satellite candidates have very high detection completeness ($>90\%$, see Table \ref{tab:sats}). We only consider satellite candidates that are within the projected \rvir in the analysis below.

\subsection{Theoretical Model}\label{sec:theory}

A primary goal of this work is to compare observed dwarf galaxy satellite systems with theoretical predictions. While cosmological hydrodynamic simulations have advanced our understanding of dwarf galaxy evolution, they face limitations in the low-mass regime. Large-box cosmological simulations \citep[e.g.,][]{Schaye2015,Springel2018} provide statistical samples but lack the resolution necessary to explore the low-mass regime properly. Zoom-in simulations achieve a much higher resolution and have advanced our understanding of satellite populations of dwarfs \citep[e.g.,][]{Munshi2019,Jahn2019}, but are limited to small sample sizes, making statistical comparisons difficult, especially for surveys like \elvesdwarf where only classical dwarfs can be detected. Additionally, numerical simulations at dwarf galaxy scales often suffer from artificial subhalo disruption \citep{vdBosch2018}, complicating the efforts to characterize satellite populations statistically.

Semi-analytical models, instead, offer the opportunity to generate a large number of satellites with high mass resolution with much less computational cost. We thus employ \satgen\footnote{\url{https://github.com/JiangFangzhou/SatGen}} \citep{Jiang2021,Green2021}, a semi-analytical model for generating satellite galaxies for host galaxies of specified mass and redshift. \satgen is computationally more efficient than simulations, allowing us to generate statistical samples of satellites. This is especially important for \elvesdwarf where the satellite abundance per host is very low. The model has successfully reproduced observed satellite statistics in Milky Way-like environments \citep{Jiang2021,Danieli2023,Monzon2024}. We briefly summarize our \satgen runs below and refer interested readers to \citet{Jiang2021} and \citet{Danieli2023} for methodological details.

In \satgen, all host halos are in isolation by construction. For a given host halo mass, \satgen generates halo merger trees and then evolves subhalos along their orbits under the tidal effects from the host halo and dynamical friction. The model also emulates baryonic processes (e.g., stellar feedback) based on hydrodynamical simulations, which is crucial as subhalos with different dark matter structures due to the baryonic effects evolve differently under tidal stripping. We use a halo response model that emulates bursty stellar feedback similar to the NIHAO simulations \citep{Tollet2016,Freundlich2020}, with a stripping efficiency of $\alpha=0.6$ following \citet{Jiang2021} and \citet{Green2021}. NIHAO-like feedback models tend to produce cored subhalos, and other feedback models might produce cuspy subhalos, which are more resistant to disruption.

Unlike the \satgen runs for Milky Way-mass hosts that typically include a disk component, we omit disk and bulge components for our dwarf hosts as they likely do not have significant stellar disks \citep{Jahn2019}. Omitting the disk component enhances the number of surviving satellites \citep{Jiang2021,Green2022}. Subhalos are considered disrupted when their mass falls below $M_{h} = 10^{7}\ M_\odot$, providing sufficient resolution for our analysis. Exploration of different halo response models, disk fraction, stripping efficiency, and resolution, is beyond the scope of this paper.

We create a library of satellite systems for dwarf hosts with halo masses spanning $10^{10.5} < M_h < 10^{12}\ M_\odot$ at $z=0$ with 0.01 dex steps in halo mass. To compare with hosts in observations, we convert the host halo mass to stellar mass using the SHMR from \citet{RP2017}. The host stellar mass range is thus $10^{7.6} < M_\star < 10^{10.5}\ M_\odot$, covering from sub-SMC to sub-MW. For each host mass, we run 11 \satgen realizations with different assembly histories. For satellites, rather than using stellar masses predicted by \satgen, we assign stellar masses to subhalos based on various SHMRs using the peak mass $M_{\rm peak}$ of subhalos. This allows us to test which SHMR matches observations better (see \S\ref{sec:nsat}). We do not assign physical sizes, surface brightness, or optical colors to the subhalos. We also do not attempt to include reionization effects, which have minimal influence on our satellite mass range $M_\star \gtrsim 10^{5.6}\ M_\odot$ \citep[e.g.,][]{Dooley2017a}.

A classical reference in this field is \citet{Dooley2017a,Dooley2017b}, who predict the satellite abundance of isolated dwarf galaxies by sampling the analytic subhalo mass functions extracted from N-body simulations and then populated subhalos using various SHMRs. Our predictions are consistent with their results within the uncertainty when using the same SHMR. 

Because of the SBF distance uncertainty (10--15\%), a small fraction of foreground and background field galaxies will be classified as satellites. \citet{Carlsten2021} found this fraction to be $\sim 10\%$ for MW-mass hosts. We do not correct for this effect in this work since it is quite small and defer it to future studies.

% NGC5023 has very small coverage in HSC, so we removed it from the list. 
% NGC4592 and NGC4765 (TF distance) are in SSP. why not include them?? Will discuss NGC4592 in results.
% Cirrus: Sandage in 70s \url{https://adsabs.harvard.edu/full/1976AJ.....81..954S}

\begin{figure*}
    \centering
    \includegraphics[width=1\linewidth]{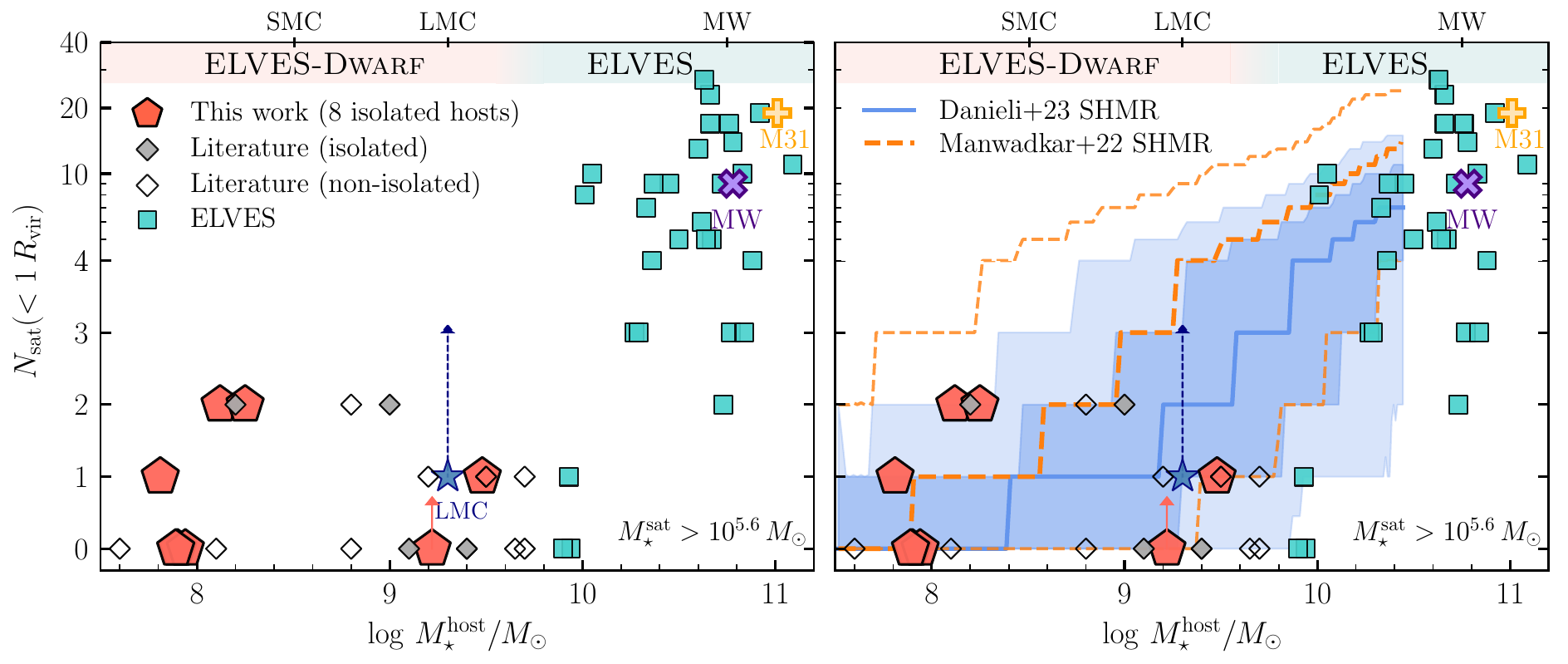}
    \caption{
    Satellite abundance of dwarf galaxies as a function of host stellar mass. \textit{Left panel:} Satellite abundances \nsat for our 8 host galaxies (red pentagons), counting only satellites with $M_\star \gtrsim 10^{5.6}\ M_\odot$ within the projected virial radius. Upward arrows indicate upper limits after accounting for the unconfirmed satellite candidates. For comparison, we include isolated hosts (filled gray diamonds) and non-isolated hosts (open diamonds) from the literature, and more massive hosts from the ELVES survey (turquoise squares). The LMC (blue star) includes the SMC and potentially Carina and Fornax as its satellites. Note that NGC~5238 and UGC05427 overlap at $\log M_\star/M_\odot \approx 7.9$ due to their similar masses. \textit{Right panel:} Theoretical predictions of \nsat from semi-analytical model \satgen using stellar-to-halo mass relations from \citet{Danieli2023} (blue) and \citet{Manwadkar2022} (orange). Solid lines show median predictions while shaded regions represent $1\sigma$ and $2\sigma$ confidence intervals. The observed satellite abundances agree well with both models within their $2\sigma$ ranges, with the \citet{Danieli2023} relation slightly favored by non-isolated hosts. We find no evidence of a ``missing satellite problem'' in the dwarf galaxy regime, particularly for the low-mass hosts ($\log M_\star/M_\odot < 8.5$) that were previously unexplored.
    }
    \label{fig:Nsat}
\end{figure*}

\begin{figure*}[htbp!]
    \centering
    \includegraphics[width=1\linewidth]{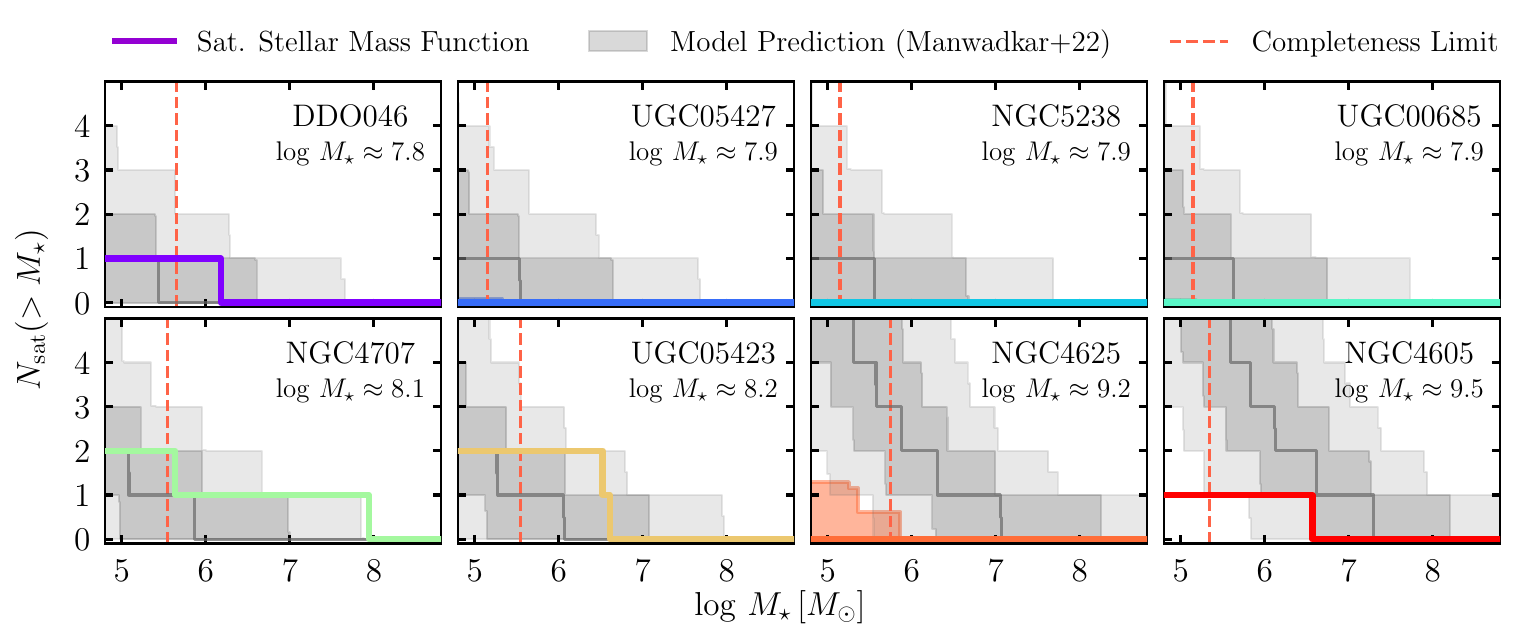}
    \caption{
    Cumulative satellite stellar mass functions for the 8 dwarf hosts in our sample, arranged by increasing host stellar mass. Only satellites within the projected virial radius are included. Solid lines show the observed mass functions, while for NGC~4625, the shaded region indicates the result when including unconfirmed satellite candidates. Theoretical predictions from \satgen using the \citet{Manwadkar2022} SHMR are shown as gray lines (median) with light and dark gray shaded regions representing $1\sigma$ (68\%) and $2\sigma$ (95\%) confidence intervals. Red vertical dashed lines mark the completeness limit for each host. The observed satellite mass functions are fully consistent with model predictions within the $2\sigma$ range.
    }
    \label{fig:sat_mass_function}
\end{figure*}

\subsection{Satellite Abundance}\label{sec:nsat}

With the SBF distances and satellite confirmation status in hand, we calculate the satellite abundance \nsat for each host. The nominal \nsat is the number of confirmed satellites. For unconfirmed satellite candidates (due to low SBF S/N or lack of HSC coverage), we assign a satellite probability $P_{\rm sat}$ based on the absolute magnitude and surface brightness (see \S\ref{sec:psat}). We also assign $P_{\rm sat}=1$ for confirmed satellites and $P_{\rm sat}=0$ for rejected satellites. Thus, the sum of these probabilities $P_{\rm sat}$ for all satellite candidates of a given host represents the upper limit of satellite abundance $N_{\rm sat,max}$ (see Table \ref{tab:sat_status}). 

The left panel of Figure \ref{fig:Nsat} shows the satellite abundances of our 8 hosts as red pentagons, with upward arrows indicating the upper limits of \nsat when including unconfirmed candidates. The arrow of NGC~4625 is the most visible because it has the most unconfirmed candidates. Note that NGC~5238 and UGC05427 overlap in Figure \ref{fig:Nsat} due to their similar stellar masses ($\log M_\star/M_\odot\approx 7.9$).

For comparison, we include satellite abundances from the literature (Appendix \ref{sec:literature}), showing isolated hosts as filled gray diamonds and non-isolated hosts (tidal index $\Theta_1 \geqslant 0$) as open diamonds. The LMC is highlighted as a blue star. We only include hosts with confirmed satellites having direct distance measurements and only count satellites more massive than our completeness limit ($M_\star \gtrsim 10^{5.6}\ M_\odot$). We caution that not all literature surveys reach this limit. We also plot the \nsat from the ELVES survey \citep{CarlstenELVES2022}, calculated for satellites within projected \rvir with $M_\star \geqslant 10^{5.6}\ M_\odot$.

Our dwarf hosts have between 0 and 2 satellites. Most of our hosts are less massive than the SMC, probing a regime rarely explored previously. Only M96-DF6, an ultra-diffuse galaxy in the M94 group, has been reported to potentially host an ultra-faint satellite in this mass range \citep{Muller2023}. Interestingly, our sub-SMC-mass hosts, all being isolate dwarfs, show considerable variation in satellite abundance: two hosts (NGC~4707 and UGC~05423) have 2 satellites each (similar to NGC~3109 from literature), DDO046 (with $\sim 1/5$ SMC's stellar mass) has one satellite, while the remaining three sub-SMC-mass hosts have zero satellites. This provides the first systematic view of satellite populations around such low-mass dwarf galaxies. Two LMC analogs (NGC~4605 and NGC~4625) have one and zero satellites, respectively, consistent with both literature results and the LMC itself.

To evaluate whether these observations align with theoretical expectations, we compare them with the predictions from \satgen (see \S\ref{sec:theory}) using two different SHMRs: \citet{Manwadkar2022} (hereafter \citetalias{Manwadkar2022}) and \citet{Danieli2023} (hereafter \citetalias{Danieli2023}). The SHMR from \citetalias{Manwadkar2022} is derived from the semi-analytical model \code{GRUMPY} \citep{Kravtsov2022}, calibrated against Local Group dwarf galaxies. We take the SHMR with a small constant scatter ($\sigma=0.06$) from \citetalias{Danieli2023}, which is based on satellites of MW analogs from the ELVES survey and has a smaller scatter than \citetalias{Manwadkar2022} and other SHMRs \citep[e.g.,][]{Garrison-Kimmel2017,Nadler2020}.

For each host stellar mass, we select \satgen runs within a 0.3 dex mass bin (roughly the uncertainty of the host stellar mass measurement), then populate subhalos using each SHMR based on the peak subhalo mass, and calculate \nsat for satellites with $M_\star > 10^{5.6}\ M_\odot$. Each mass bin includes 330 \satgen runs. The results are shown in the right panel of Figure \ref{fig:Nsat}. The solid blue line represents the median \nsat predicted using the SHMR from \citetalias{Danieli2023} with blue shaded regions showing $1\sigma$ (68\%) and $2\sigma$ (95\%) confidence intervals. Orange lines show the predictions using the SHMR from \citetalias{Manwadkar2022}. \citetalias{Danieli2023} predicts lower \nsat values than \citetalias{Manwadkar2022}, though they remain consistent within $1\sigma$. This difference is because the SHMR from \citetalias{Manwadkar2022} is slightly higher than the one from \citetalias{Danieli2023} and has a substantially larger scatter ($\sigma \approx 0.2$~dex versus $\sigma \approx 0.06$~dex). 

The observed satellite abundance of our hosts is fully consistent with both models within the $2\sigma$ confidence level. The same conclusion is found for isolated hosts from the literature. Interestingly, we find that a few non-isolated hosts from the literature (open diamonds) around $M_\star \sim 10^{9.7}\ M_\odot$ have lower \nsat values and are inconsistent with the prediction using the SHMR from \citetalias{Manwadkar2022} at a $2\sigma$ level. This potentially suggests that the SHMR of satellites depends on the environment. 

Recently, the ID-MAGE survey \citep{Hunter2025} found a satellite abundance of $N_{\rm sat}=4.0 \pm 1.4$ for LMC-like hosts with a similar detection limit as our survey. This \nsat is higher than our result but is still consistent with \satgen predictions. Since their candidates do not have direct distance information, \nsat is estimated by statistically subtracting the contribution from background galaxies and assuming a specific satellite radial distribution. Their \nsat is thus likely to be subject to considerable uncertainty.

\subsection{Satellite Stellar Mass Function}
Figure \ref{fig:sat_mass_function} shows the cumulative satellite stellar mass functions for each host, only including the satellite candidates within the projected \rvir. For NGC~4625, the shaded region indicates the cumulative number when including the $P_{\rm sat}$ of the unconfirmed satellite candidates. To make theoretical predictions, for each host, we select \satgen runs within a 0.3 dex mass bin and populate subhalos using the \citetalias{Manwadkar2022} SHMR. The predicted satellite mass functions are shown as gray solid lines (median) with gray shaded regions showing $1\sigma$ and $2\sigma$ ranges. Vertical dashed lines mark the completeness limit for each host, and the hosts are arranged in order of increasing stellar mass.

The observed satellite mass functions agree well with theoretical predictions for hosts in our sample. DDO046, UGC05423, and NGC~4605 show particularly good agreement with the model. For a few sub-SMC mass hosts (UGC05427, NGC~5238, and UGC00685), the absence of satellites above our detection completeness limit is fully consistent with theoretical expectations. NGC~4707 exhibits a slightly more massive and abundant satellite system than average but remains within the $2\sigma$ prediction range. NGC~4625 shows fewer satellites than expected, likely because it is the most distant host in our sample and thus has poorer completeness. Nevertheless, its satellite mass function remains consistent with the model within $2\sigma$ up to our completeness limit. The discrepancy is further reduced when accounting for unconfirmed satellite candidates. 

We also compare the observed satellite mass function with the predictions using the \citetalias{Danieli2023} SHMR. Because it is lower than the \citetalias{Manwadkar2022} SHMR and has a much smaller scatter, it predicts a slightly smaller cumulative number of satellites at a given satellite mass. For this reason, the observed satellite mass functions of NGC4707 and UGC05423 are higher than the $2\sigma$ upper limit of the \citetalias{Danieli2023} prediction, being slightly inconsistent with the model prediction. We do not attempt to stack the individual satellite mass functions because of the small sample size and the varying completeness limit.

\begin{figure}[htbp!]
    \centering
    \includegraphics[width=1\linewidth]{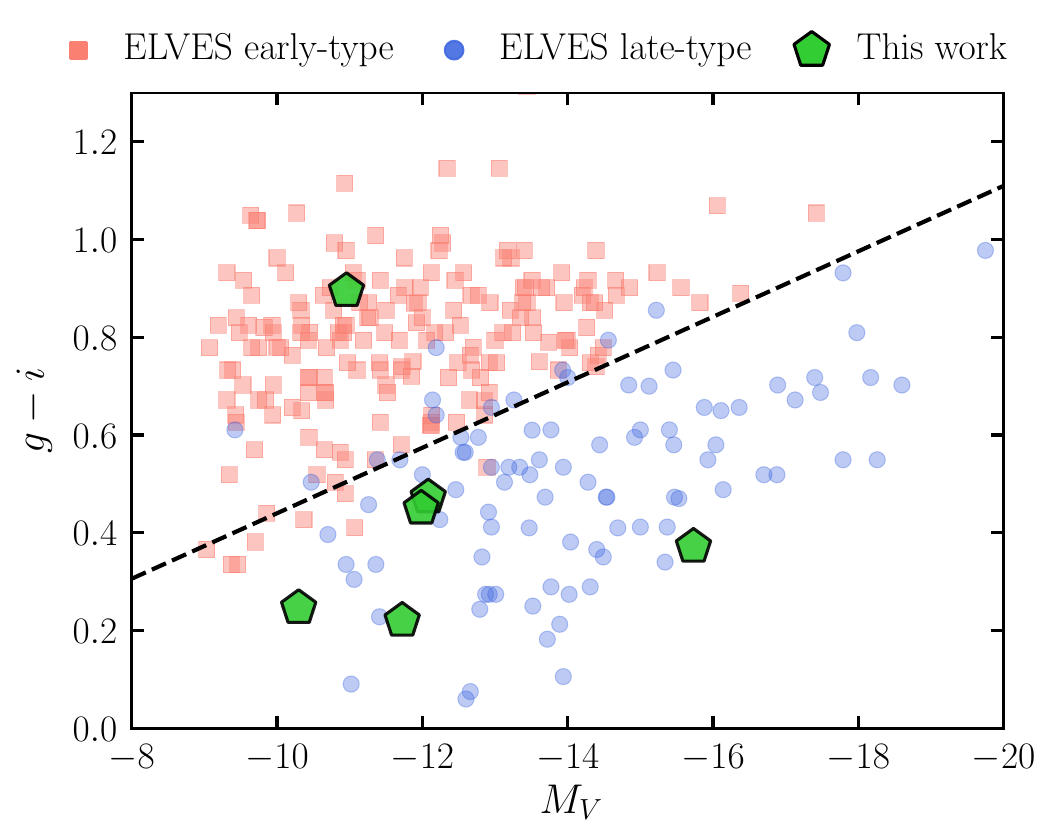}
    \caption{
    Color-magnitude diagram for the 6 confirmed satellites in this work (green pentagons) and the confirmed satellites of MW analogs (squares) from the ELVES survey \citep{CarlstenELVES2022}. The ELVES satellites are visually classified into early-type and late-type based on their morphology. The two classes can be nicely separated by the black dashed line proposed by \citet{ELVES-I}. Most of our confirmed satellites are blue and star-forming, implying a very low quenched fraction at all satellite masses.
    }
    \label{fig:sats_color}
\end{figure}

\subsection{Satellite Quenching}

The star formation properties of satellites provide crucial insights into their evolution and the environmental effects such as tidal and ram-pressure stripping \citep{GunnGott1972}. Among six confirmed satellites, five are star-forming dwarfs showing blue optical colors and late-type morphologies. Only one satellite (dw1245p6158, associated with NGC~4605) shows an early-type morphology with a red color. 

Figure \ref{fig:sats_color} shows the color and absolute magnitude of our confirmed satellites (green pentagons), together with the satellites of MW analogs from the ELVES survey \citep{CarlstenELVES2022}. We convert the $g-r$ color to $g-i$ color using $g-i = 1.53 \, (g-r) - 0.032$ from \citet{ELVES-I} which is derived from the simple stellar population models. The ELVES satellites are classified into early-type (red and smooth, shown as red squares) and late-type (blue, asymmetric, and clumpy, shown as blue dots) that can be well-separated by a linear relation $g-i = -0.067\, M_V - 0.23$ \citep{ELVES-I}. As also shown in \citet{Font2022,SAGA-III,Li2023}, this mass-dependent color cut works well on separating star-forming satellites from quiescent ones. Figure \ref{fig:sats_color} indicates a remarkably low quenched fraction for satellites of dwarfs, as most satellites we confirmed are star-forming. 

Our finding contradicts the LBT-SONG survey \citep{Davis2024}, which found no star-forming satellites in their sample. We note that their hosts are more massive than ours and include both isolated and non-isolated hosts. Direct distances are also needed to confirm their candidates as real satellites. Similarly, using the FIRE simulations, \citet{Jahn2022} found that the satellite quenched fraction around LMC analogs is comparable to that of MW analogs. Compared to satellites of MW analogs, satellites of dwarfs typically have earlier infall times, but the less dense environment may cause a longer quenching timescale due to reduced ram-pressure stripping. The net effect depends on which factor dominates. Our observed low quenched fraction suggests a longer quenching timescale than for MW satellites. This has interesting implications for quenching mechanisms in lower-density environments.

\section{Discussion and Summary}\label{sec:discussion}

In this work, we present the first results from the \elvesdwarf survey, which aims to construct a statistical sample of satellite systems of isolated dwarf galaxies. We focus exclusively on isolated dwarfs to avoid ambiguity when associating satellite candidates with their hosts, and to make cleaner comparisons with theoretical models by excluding the additional tidal effects and ram-pressure stripping from more massive galaxies on satellites. While our host selection differs from several other surveys that include hosts in all environments, comparing satellites of isolated and non-isolated hosts could also be interesting and could reveal potential environmental dependence in SHMR.

We systematically searched for satellite candidates around eight dwarf hosts within 12 Mpc, spanning a mass range from LMC analogs to sub-SMC dwarfs. Using the surface brightness fluctuation technique to measure distances, we confirmed six satellites above our completeness limit $M_\star > 10^{5.6}\, M_\odot$, as summarized in Table \ref{tab:sat_status}. Notably, four of our dwarf hosts have no confirmed satellites, while the remaining four have either one or two confirmed satellites each. As shown in Figure \ref{fig:Nsat}, when combined with the ELVES results, we find that more massive hosts tend to have more satellites, but the trend weakens at lower masses. The satellite abundance \nsat and the satellite stellar mass function show substantial host-to-host scatter. The predictions from \satgen are overall consistent with the observation, and we find no evidence of a ``missing satellite'' problem in the dwarf galaxy regime. 

The abundance of classical satellites of isolated dwarf galaxies might have strong implications for constraining the SHMR. In Figure \ref{fig:Nsat}, we compare our observed \nsat with the predictions using two different SHMRs from \citetalias{Manwadkar2022} and \citetalias{Danieli2023}. While both models describe the sample well, we find some evidence that the \citetalias{Danieli2023} model better agrees with the $N_{\rm sat}$ of non-isolated dwarf hosts, whereas the \citetalias{Manwadkar2022} model better describes both $N_{\rm sat}$ and the satellite mass function of isolated hosts. However, this appears to contradict \citet{Christensen2024}, who found that dwarf galaxies in more isolated environments have lower stellar masses for a given peak halo mass. We caution that many non-isolated hosts from the literature were not searched out to $R_{\rm vir}$ and have heterogeneous detection limits. Accounting for detailed selection functions will be critical to fully disentangle any environmental dependencies in the SHMR. Also, as emphasized by \citet{Monzon2024}, a large sample of dwarf hosts is needed to constrain the slope and scatter of the SHMR.

The number of classical satellites may also inform our expectations regarding ultra-faint satellites. \citet{Santos-Santos2022} demonstrated that a scale-free `power-law' SHMR (similar to \citetalias{Danieli2023}) and an SHMR with a sharp cut-off at very low mass \citep[e.g.,][]{Fattahi2018} predict distinguishable numbers of classical satellites of dwarf galaxies, but predict dramatically different number of ultra-faint dwarfs. By constraining the SHMR using classical satellites, we can make predictions about ultra-faint populations of nearby dwarf galaxies that can be tested by future observations. 

We find a remarkably low quenched fraction for our confirmed satellites; only one out of six is quenched. Although the quenched fraction of dwarfs in the field \citep{Geha2012} and in massive groups has been thoroughly studied \citep[e.g.,][]{Wetzel2015,Font2022,Pan2022,Li2023,Greene2023,SAGA-IV}, very little is known about the quenching of satellites in less dense environment. For dwarf hosts, their lower-density circumgalactic media might lead to low ram-pressure stripping efficiency. Comparing the quenched fraction in different environments would help break degeneracies between the infall time distribution and the quenching timescale.

A larger sample of dwarf hosts is needed to fully sample the satellite systems, better understand the SHMR, probe environmental effects, and determine the quenched fraction. In Paper II (Li et al., in prep.) of the \elvesdwarf series, we will present an additional $\sim15$ systems and provide a more comprehensive analysis of satellite properties. Looking ahead, the Vera C. Rubin Observatory's Legacy Survey of Space and Time \citep{LSST2019}, the \textit{Euclid} mission \citep{Euclid_Overview,Euclid-Q1}, and the Nancy Grace Roman Space Telescope \citep{Spergel2015,Akeson2019} will enable us to probe the satellite mass function below our current completeness limits \citep[e.g.,][]{Simon2019,Mutlu-Pakdil2021} and shed lights on galaxy formation in the low-mass regime.

\appendix

\section{Detection completeness of individual hosts}\label{sec:completeness}

\begin{figure*}
    \centering
    \includegraphics[width=1\linewidth]{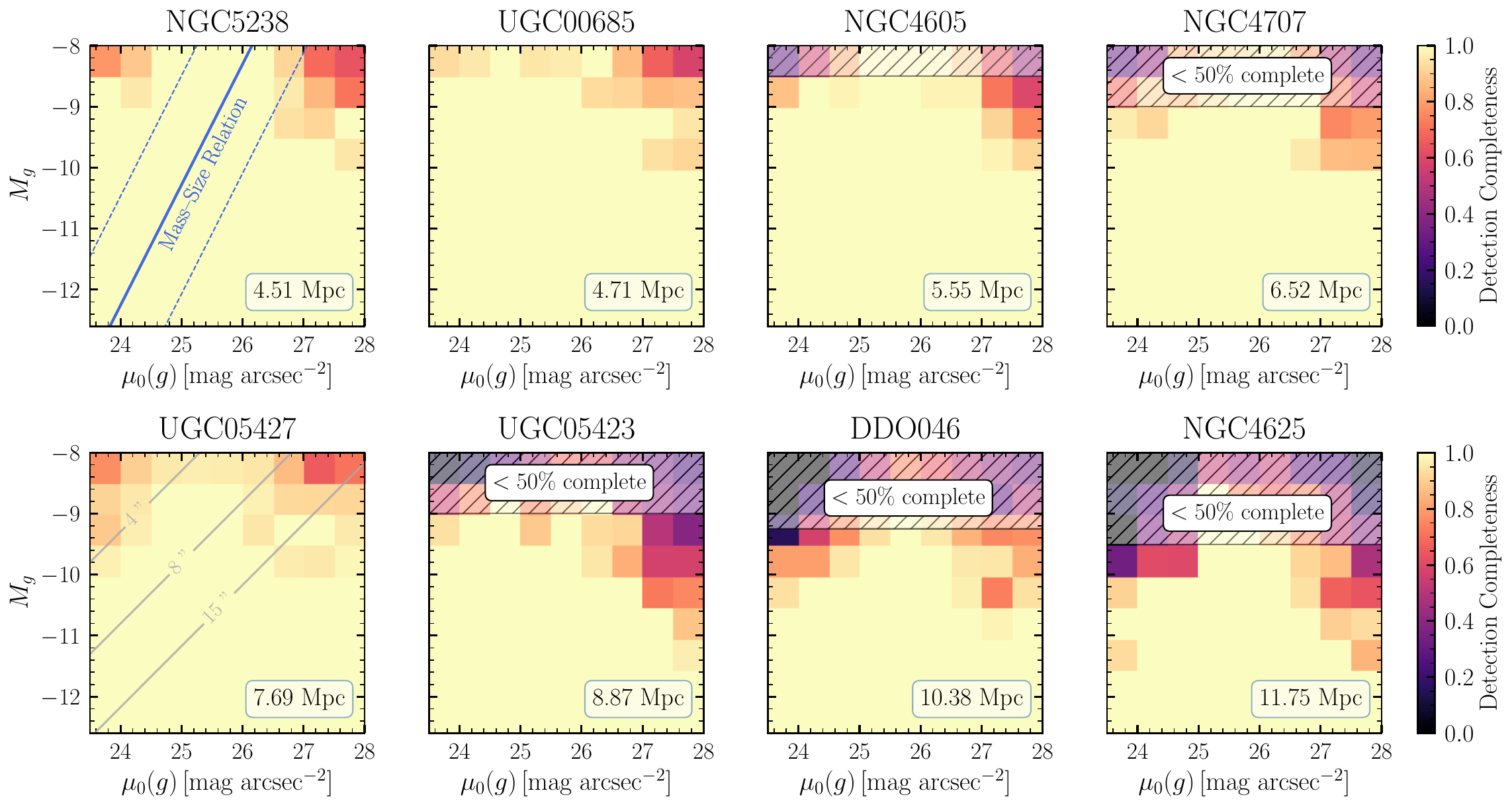}
    \caption{Detection completeness of each host as a function of central surface brightness $\mu_0(g)$ and absolute magnitude $M_g$, arranged by increasing distance. The white-shaded region shows where the completeness drops below 50\%.}
    \label{fig:completeness}
\end{figure*}

We characterize our detection completeness by injecting mock dwarf galaxies into the Legacy Surveys data and recovering them with the detection pipeline. The mock galaxies are designed to cover a wide range in total magnitude, surface brightness, and color to mimic the real satellite population. The mock galaxies follow a single-\sersic model with $n=1$. We uniformly sample the absolute $g$-band magnitude in $-12.5 < M_g < -7.5$~mag, central $g$-band surface brightness in $23.5 < \mu_0(g) < 28.0\ \sbunit$, color in $0.3 < g-r < 0.9$, and ellipticity in $0 < \varepsilon < 0.5$. The properties of these mock galaxies roughly follow the real satellite population in the ELVES survey \citep{ELVES-I}. For each brick of Legacy Surveys data, we randomly inject 12 mock galaxies to the coadd images at a time, run them through our detection pipeline, and match the detected objects with the input catalog. We repeat injecting mock galaxies several times until we have enough statistics to calculate the detection completeness, defined as the number of detected objects divided by the number of injected objects. We find that the detection completeness mainly depends on the total magnitude and central surface brightness, with little dependence on color and ellipticity. Mock galaxies that are masked by the bright star mask are not counted when calculating the completeness. The area lost to the star masks is considered separately by the unmasked fraction in Table \ref{tab:hosts}.

Figure \ref{fig:completeness} shows the detection completeness as a function of absolute magnitude $M_g$ and central surface brightness $\mu_0(g)$. There are at least 15 mock galaxies in each surface brightness and absolute magnitude bin. To better guide the readers' eyes, we show the constant angular size lines in gray. The detection completeness is low for both small and large objects with low surface brightness. The completeness decreases as a function of distance and varies from host to host because of different data quality and depth. For hosts at $D<8$~Mpc, we are $>50\%$ complete to $M_g \approx -8.5$. However, for the three hosts at $D>8$~Mpc, our $50\%$ completeness limit is about $M_g \approx -9.5$. The white-shaded region in Figure \ref{fig:completeness} shows where the completeness drops below 50\% for each host.  

To place the detection completeness in context, we also show the average mass-size relation (blue line) and its $1\sigma$ scatter (blue dashed lines) assuming a Sersic index of $n=1$ and $g-r=0.4$. The detection completeness is quite uniform along the mass-size relation, except for the very faint end where the completeness drops below $50\%$. 

We also cross-validate our completeness by matching our detections with the Systematically Measuring Ultra-diffuse Galaxies (SMUDGes) survey \citep{Zaritsky2019,Zaritsky2023}. Every object that is in the SMUDGes catalog is detected by us. However, we removed several of the SMUDGes objects after visually inspecting the HSC data because they appear to be background galaxies with spiral arms of bulges.

\section{\lowercase{$r$}-band SBF calibration}\label{sec:sbf_calib_r}
Despite the fact that optical SBF is usually performed in the $i$-band \citep[e.g.,][]{Tonry1988,Cantiello2018,Carlsten2019,Kim2021} because of better seeing and more prominent SBF signal \citep{Greco2021}, it is still possible to measure SBF in other optical bands. \citet{Carlsten2021} presented an SBF calibration in the $r$-band, by converting the $i$-band calibration in \citet{Carlsten2019} to the $r$-band using filter conversions. They used simple stellar populations (SSPs) with ages between 3 and 10 Gyr and metallicity $-2 < \mathrm{[Fe/H]} < 0$ to convert $(g-i)$ color to $(g-r)$ color, and convert $i$-band SBF magnitude ($\overline{M}_{i}$) to $r$-band ($\overline{M}_{r}$). This $r$-band calibration simply inherits the scatter from the $i$-band calibration in \citet{Carlsten2019}. 

However, the scatter of the $r$-band calibration is likely underestimated in \citet{Carlsten2019}. While the color conversion has a quite small scatter ($<0.02$ mag), the SBF magnitude conversion has a non-negligible scatter ($\sim 0.1$~mag, see Fig. 11 in \citealt{Carlsten2021}) and should also contribute to the scatter of the $r$-band SBF calibration. 

To better characterize the scatter, we use a sample of 14 dwarf galaxies in \citet{Carlsten2021} to re-calibrate the SBF--color relation in the $r$-band. These galaxies have SBF measurements in CFHT $r$-band and TRGB distances (see Table 1 and Figure 2 in \citealt{Carlsten2021}). We then use the Bayesian algorithm \code{LINMIX} \citep{linmix} to fit a linear relation $\overline{M}_{r,\ \rm CFHT} = \alpha_0 + \beta_0 \cdot (g-r)_{\rm CFHT}$. We assume the measurement uncertainties in colors and SBF magnitudes are independent and Gaussian, and the priors on intercept and slope are uniform. We note that $\alpha_0$ and $\beta_0$ are highly covariant, and their posterior distribution can be well-approximated by a multivariant Gaussian distribution with the following parameters:
\vspace{0em}
\begin{equation}
\begin{aligned}
\text{Mean:} \quad \mathbf{\mu_{\rm CFHT}} &= \begin{pmatrix} -2.753 & 3.638 \end{pmatrix} \\
\text{Covariance:} \quad \mathbf{\Sigma_{\rm CFHT}} &= \begin{pmatrix}
0.3804 & -0.8507 \\
-0.8507 & 1.9705
\end{pmatrix}
\end{aligned}
\end{equation}

\begin{figure}
    \centering
    \includegraphics[width=0.9\linewidth]{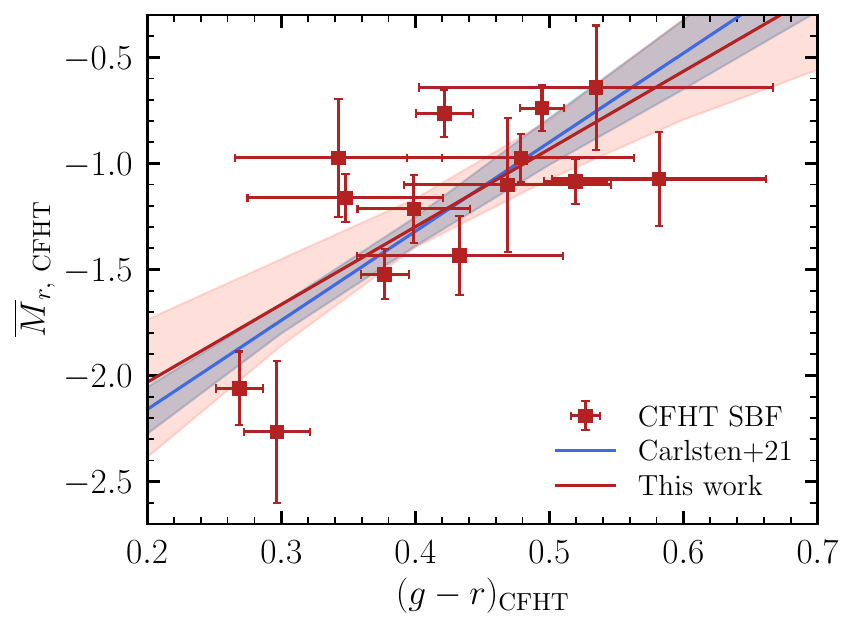}
    \caption{The re-calibrated $r$-band SBF--color relation using the sample from \citet{Carlsten2021}. The newly calibrated relation has a similar slope to \citet{Carlsten2021} but has a larger scatter.}
    \label{fig:SBF_calib_r}
\end{figure}

This new calibration (red line) is illustrated in Figure \ref{fig:SBF_calib_r}, together with the calibrator galaxies (red data points) and the calibration (blue line) in \citet{Carlsten2021}. The shaded regions correspond to the 68\% ($1\sigma$) credible intervals. The new calibration is slightly shallower than the original calibration and has a larger scatter. As an example, at $(g-r)_{\rm CFHT} = 0.5$, the $1\sigma$ scatter in $M_{r}$ is 0.11~mag in the original calibration, and is 0.15~mag in this new calibration. 

We further convert this calibration from the CFHT filter system to the HSC one. We generate SSPs with $3 < \mathrm{age} < 10$~Gyr, $-2 < \mathrm{[Fe/H]} < 0$ using the MIST isochrones \citep{Choi2016} for both filter systems, then fit linear relations between colors and SBF magnitudes: $(g-r)_{\rm CFHT} = a_1 + b_1 \cdot (g-r)_{\rm HSC}$; $\overline{M}_{r,\rm HSC} = a_2 + b_2 \cdot M_{r, \rm CFHT}$. The best-fit values of these parameters are $a_1 = 0.0077,\ b_1 = -0.049,\ a_2 =0.024,\ b2=-0.0031$. The resulting $r$-band calibration in the HSC filter system is thus:
$$\overline{M}_{r,\rm HSC} = \alpha + \beta \cdot (g-r)_{\rm HSC},$$
where $\alpha = a_2 + b2\,(\alpha_0 + \beta_0 a_1)$ and $\beta = \beta_0 b_1 b_2$ are the intercept and slope of the SBF calibration in the HSC filter system. The joint distribution of $(\alpha, \beta)$ can also be approximated by a multivariant Gaussian with:
\vspace{0em}
\begin{equation}
\begin{aligned}
\text{Mean:} \quad \mathbf{\mu_{\rm HSC}} &= \begin{pmatrix} -2.7037 & 3.4753 \end{pmatrix} \\
\text{Covariance:} \quad \mathbf{\Sigma_{\rm HSC}} &= \begin{pmatrix}
0.3621 & -0.7729 \\
-0.7729 & 1.7018
\end{pmatrix}
\end{aligned}
\end{equation}
The difference between HSC and CFHT calibration is small ($\Delta \overline{M}_r \sim 0.05$~mag for $g-r=0.5$). In future works, we will use the whole ELVES and \elvesdwarf{} satellite sample to better calibrate the $r$-band SBF--color relation.

\begin{figure}[htbp!]
    \centering
    \includegraphics[width=0.8\linewidth]{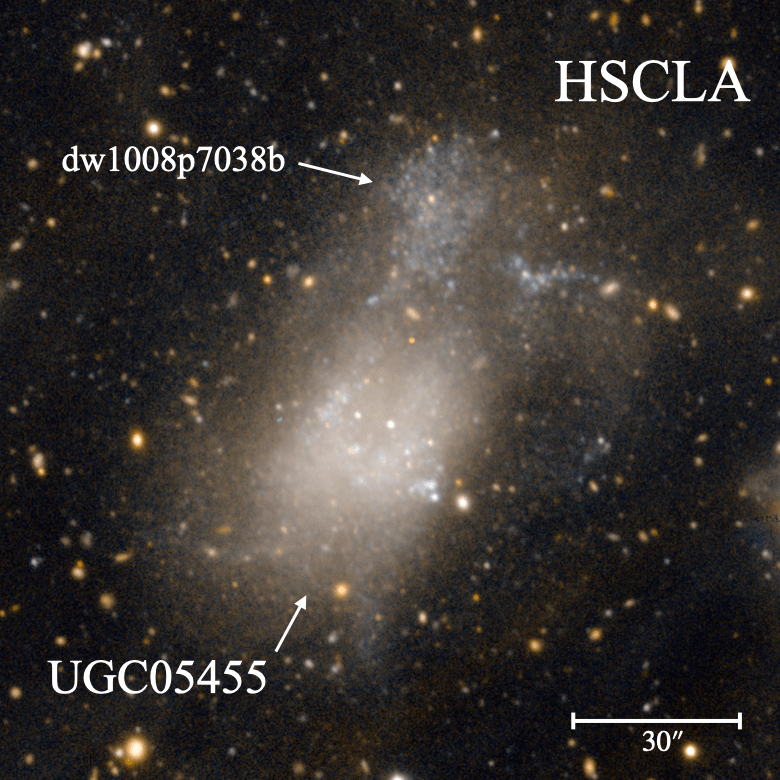}
    \caption{Color-composite image of two satellite candidates (UGC05455 and dw1008p7038b) of UGC05423. We use the HSCLA $g$-band image for the blue channel, $i$-band image for the red channel, and $(g+i)/2$ for the green channel. It is clear that dw1008p7038b has a bluer color and shows stronger SBF signals, making it a foreground galaxy that happens to overlap with the background galaxy UGC05455 in the sky.}
    \label{fig:UGC05455}
\end{figure}

% \begin{figure*}[htbp!]
%     \centering
%     \includegraphics[width=1\linewidth]{figures/gallery.png}
%     \caption{Cutout images of all confirmed satellites. The top panels show the color-composite images from the Legacy Surveys DR10, and the bottom panels show the single-band data from HSC. These images have 1\arcmin{} on a side and are stretched to show the SBF signals.}
%     \label{fig:stamps}
% \end{figure*}

% \section{HSC cutout images of confirmed satellites}\label{ap:stamps}

\section{HSC image of two satellite candidates in the UGC05423 field}\label{ap:UGC05455}

Out of all confirmed satellites, dw1008p7038b is an intriguing system due to its overlap with a background galaxy UGC05455. Figure \ref{fig:UGC05455} shows the color-composite image of this system. This color image is made from the HSCLA data in the $g$ and $i$ bands. We use the $g$-band image for the blue channel, the $i$-band image for the red channel, and $(g+i)/2$ for the green channel. It is clear in this image that dw1008p7038b is bluer than UGC05455 and shows stronger SBF signals. 

The measured SBF distance of UGC05455 is $D_{\rm SBF} = 14.62^{+1.54}_{-1.42}\ \rm Mpc$, putting it to be a background galaxy. We then measure the SBF distance of dw1008p7038b to be $D_{\rm SBF} = 10.63^{+1.49}_{-1.16}\ \rm Mpc$. Because of the overlap between dw1008p7038b and UGC05455, the outskirts of UGC05455 will contribute to the smooth galaxy model of dw1008p7038b and bias the measured SBF distance to higher values. Yet, the $2\sigma$ range of the measured SBF distance is still consistent with the host distance. Therefore, we consider dw1008p7038b a real satellite of UGC05423. Spectroscopic observations can easily test whether it has a similar velocity as the host.

\section{Satellite Distances and Photometric Properties}
Table \ref{tab:sats} presents the SBF distance measurements and the photometric properties of all satellite candidates. 

\include{sat_cat}

\section{Literature results}\label{sec:literature}
In this appendix, we compile results from the literature on satellite systems of dwarf galaxies. The majority of the literature is from the MADCASH and the LBT-SONG surveys. Given the detection limit of most surveys, we only include ``classical'' satellites with $M_\star > 10^{5}\,M_\odot$, omitting the ultra-faint dwarfs. We also limit to host galaxies with $M_\star < 10^{10}\ M_\odot$, as higher-mass hosts are searched in various other surveys, including ELVES \citep{CarlstenELVES2022}. We do not include the DDO068 system because this system is undergoing a merger, making it challenging to characterize the properties of the satellite candidates \citep{Tikhonov2014,Annibali2019,Annibali2023}. Several latest surveys, such as LIGHTS \citep{Zaritsky2024_LIGHTS} and ID-MAGE \citep{Hunter2025}, are not included here because most of their candidates are still unconfirmed.

Table \ref{tab:literature} presents the key properties of host dwarf galaxies and their associated satellites, including distance, $V$-band absolute magnitude $M_V$, stellar mass $M_\star$, effective radius $r_{\rm eff}$, and tidal indices (only for the hosts). The host galaxies span a wide range in mass and environment. We caution that the literature searches are not homogeneous in depth and survey footprint. Some surveys only cover a small fraction of the projected virial radius. 

This compilation serves as a useful reference for comparing our findings with previously studied systems and testing theoretical predictions against observation. It also highlights the diversity of dwarf galaxy satellites in the Local Volume. We have made extensive use of the Updated Nearby Galaxy Catalog \citepalias{Karachentsev2013} and the Local Volume Database \citep{Pace2024} when compiling this catalog. Typical uncertainties of $M_V$ and $M_\star$ are $\sim 0.2$~mag and $\sim 0.2$~dex, respectively. Each system is described individually below.

\vspace{1em}

\textit{LMC} --- The Large Magellanic Cloud (LMC) is a satellite of the Milky Way located at a distance of $D\approx 50\ \rm kpc$ with a stellar mass of $\log M_\star/M_\odot \approx 9.3$ \citep{Skibba2012LMC}. Higher and lower stellar mass estimates also exist in the literature \citep[e.g.,][]{McConnachie2012,vanderMarel2006LMC,Pace2024}. We conservatively identify only one classical satellite of the LMC: the Small Magellanic Cloud \citep[SMC;][]{Sales2011,Santos-Santos2021}. However, recent dynamical studies suggest that Carina ($\log M_\star/M_\odot \approx 6$) and Fornax ($\log M_\star/M_\odot \approx 7.6$) might also be associated with LMC \citep[e.g.,][]{Pardy2020}. The properties of SMC are taken from \citet{Pace2024}.

\textit{NGC~6822} --- NGC~6822 is at $D= 0.459\ \rm Mpc$ \citep{McConnachie2012} with a stellar mass of $\log M_\star / M_\odot \approx 8.1$. The stellar mass is converted from the $K_s$-band luminosity in \citet{Karachentsev2013} using a mass-to-light ratio of $M_\star / L_{\rm K_s} = 0.6$. \citet{Zhang2021_NGC6822} performed a dedicated satellite search but found no satellite candidates brighter than $M_V = -5$ within their survey footprint. 

\textit{M33} --- M33 (the Triangulum Galaxy) is a satellite of M31 located at a distance of $D= 0.84\ \rm Mpc$ \citep{Breuval2023} and a stellar mass of $\log M_\star / M_\odot \approx 9.7$ \citep{Corbelli2014}. It harbors no classical satellites but might be associated with three ultra-faint dwarfs \citep[][]{Martin2016M31,Martinez-Delgado2022,Collins2024,Ogami2024}.

\textit{NGC~3109} --- NGC~3109 is a dwarf galaxy at the edge of the Local Group ($D= 1.34\ \rm Mpc$, \citealt{Karachentsev2013}). It has a $V$-band absolute magnitude of $M_V \approx -15$ \citep{Sand2015} and a stellar mass of $\log M_\star/M_\odot \approx 8.2$ (derived from its $K_s$-band luminosity). Antlia, a dwarf galaxy at $D\approx 1.35\ \rm Mpc$ with $\log M_\star/M_\odot \approx 6.4$ \citep{McConnachie2012}, is associated with NGC~3109. The MADCASH survey discovered a new satellite, Antlia B, at $D=\approx 1.29\ \rm Mpc$ with $M_V\approx -9.7$ and $\log M_\star/M_\odot \approx 6.15$ \citep{Sand2015,Hargis2019}. While Sextans A, Sextans B \citep{Bellazzini2014}, and Leo P \citep{McQuinn2015} are considered as members of this loose group, they are not included in Table \ref{tab:literature} because they are beyond the project virial radius of NGC~3109. 

\textit{NGC~300} --- NGC~300 is located at $D = 2.0\ \rm Mpc$ \citep{Dalcanton2009} with a stellar mass of $\log M_\star / M_\odot \approx 9.2$ (derived from its $K_s$-band luminosity). \citet{Sand2024} discovered a satellite, Sculptor~C, at $D=2.04\ \rm Mpc$ with $M_V\approx -9.1$. 

\textit{NGC~55} --- NGC~55 is an LMC analog at $D= 2.34\ \rm Mpc$ \citep{Kudritzki2016} with a stellar mass of $\log M_\star/M_\odot \approx 9.65$ \citep{Pace2024}. Recently, \citet{McNanna2024} discovered NGC~55-dw1, a very faint satellite of NGC~55, which is fully resolved into stars in the DELVE survey \citep{DELVE2021}. This satellites has a $V$-band absolute magnitude of $M_V\approx -8.0$ and a stellar mass of $\log M_\star/M_\odot \approx 5.15$, placing it at the borderline of being an ultra-faint dwarf. NGC~55-dw1 also has an unexpectedly large size, making it the most diffuse dwarf at this mass. 

\textit{NGC~2403} --- NGC~2403 is located at $D=3.19\ \rm Mpc$ \citep{Karachentsev2013} with $M_V\approx -19.1$ and $\log M_\star/M_\odot = 9.7$ (derived from the $K_s$-band luminosity). The MADCASH survey has comprehensively searched for its satellites down to $M_V\approx -7.5$, and identified two confirmed satellites: DDO~44 \citep{Karachentsev1999DDO044,Carlin2019} and MADCASH-1 \citep{Carlin2016,Carlin2024}.

\textit{NGC~4214} --- NGC~4214 is at $D=3.04\ \rm Mpc$ \citep{Dalcanton2009} with $M_V\approx -17.5$ and $\log M_\star/M_\odot \approx 8.8$ (from $K_s$-band luminosity). It has a confirmed satellite, DDO~113, at $D=2.95\ \rm Mpc$ \citep{Dalcanton2009,Weisz2011} with $M_V\approx -12.2$ \citep{Garling2020}. The MADCASH survey discovered another satellite in the system, MADCASH-2 \citep{Carlin2021}, with a distance of $D=3.00\ \rm Mpc$ and $M_V=-9.15$.

\textit{NGC~4449} --- NGC~4449 is an LMC analog at $D=4.02\ \rm Mpc$ \citep{Sabbi2018}. \citet{Martinez-Delgado2012} discovered LV~J1228+4358, a satellite in this system located at $D=4.07\ \rm Mpc$ \citep{Karachentsev2013}. This satellite was also discovered by the LBT-SONG survey \citep{Garling2021}. 

\textit{NGC~3077, IC~2574, and NGC~4236} --- These three host galaxies were searched in \citet{Garling2021} with no satellite detected. It is important to note that their survey footprint covers only a small fraction of the projected virial radius of each host. Therefore, these hosts might still have undetected satellites outside the survey footprint of \citet{Garling2021}.

\textit{NGC~1156} --- This is a dwarf galaxy at $D=6.87\ \rm Mpc$ \citep{Karachentsev2013} with $\log M_\star/M_\odot \approx 9.0$ (derived from $K_s$-band luminosity). It is an isolated galaxy according to the tidal indices. \citet{Karachentsev2015_NGC1156} reported the discovery of two low-surface-brightness galaxies around NGC~1156, namely NGC~1156-dw1 and NGC~1156-dw2, with no distance measurement. \citet{Carlsten2020} discovered another satellite candidates, dw0301+2446. Another satellite candidate, named AGES J030039+254656, was found by \citet{Minchin2010} in HI surveys. None of these satellite candidates are confirmed.

\textit{NGC~5023} ---  NGC~5023 is at $D=6.09\ \rm Mpc$ with $\log M_\star/M_\odot \approx 8.8$ (derived from $K_s$-band luminosity). Its tidal indices indicate that it is in isolation. \citet{Carlsten2020} discovered two satellite candidates, dw1310+4358 and dw1314+4420, with no distance measurements.

\textit{NGC~45} --- This is a galaxy at $D=6.64\ \rm Mpc$ \citep{Karachentsev2013} with $\log M_\star/M_\odot \approx 9.1$ (derived from $K_s$-band luminosity). \citet{Muller2020} found no satellites for this host.

\textit{NGC~24} --- This is a galaxy at $D=7.31\ \rm Mpc$ \citep{Karachentsev2013} with $\log M_\star/M_\odot \approx 9.2$ (derived from $K_s$-band luminosity). \citet{Muller2020} discovered two satellite candidates, dw0009-25 and dw0010-25, around this host. No direct distances are available for these satellite candidates.

\textit{NGC~672} --- Located at $D=7.18\ \rm Mpc$ \citep{Karachentsev2013}, this galaxy has a bright satellite, IC~1727, with a TRGB distance of 7.45 Mpc \citep{Tully2013}. The host galaxy NGC~672 has a stellar mass of $\log M_\star/M_\odot \approx 9.5$ (derived from $K_s$-band luminosity), and its brightest satellite IC~1727 has a stellar mass of $\log M_\star/M_\odot \approx 9.3$. This system also has another two satellite candidates, NGC~6720-dwA and NGC672-dwD, discovered by the LBT-SONG survey \citep{Davis2024}. The latter would be an ultra-faint dwarf if confirmed. 

\textit{NGC~925} --- This galaxy is at $D=9.55\ \rm Mpc$ \citep{Karachentsev2013} and has a stellar mass of $\log M_\star/M_\odot \approx 9.9$ (derived from $K_s$-band luminosity). \citet{Davis2024} discovered a satellite candidate, d0226+3325. It has no direct distance measurement yet but shows SBF signals in its LBT images.

\textit{M96-DF6} --- This is an ultra-diffuse galaxy associated with M96 at $D=10.20\ \rm Mpc$ \citep{Cohen2018}. Recently, \citet{Muller2023} discovered dw1046+1244, a possible satellite associated with M96-DF6. This satellite candidate is resolved into single stars in HST data and has a stellar mass of $\log M_\star/M_\odot \approx 5.4$ if confirmed.

\textit{NGC~4592} --- NGC~4592 is an LMC-mass dwarf galaxy ($M_r=-18.23$~mag) at $D=9.02\ \rm Mpc$ \citep{Kim2020}. It forms a dwarf galaxy pair with NGC~4517 (also known as NGC~4437, $M_r=-20.71$~mag at $D=9.28\ \rm Mpc$). Therefore, NGC~4592 is not isolated. \citet{Kim2022} searched satellites in the group using HSC-SSP data, and found NGC~4517-dw15 and NGC~4517-dw16 within the projected virial radius of NGC~4592. These two dwarfs are confirmed to be associated with the hosts using SBF distances. Another satellite in this group, NGC~4517-dw12, is at the edge of the projected virial radius of NGC~4592, and is thus not included in Table \ref{tab:literature}. These satellites are also confirmed by the ELVES survey \citep{CarlstenELVES2022}.

\input{literature_hosts}

% \section{Group association using SBF distances}

% The long lingering question is whether 15\% accuracy of SBF distance is enough to associate satellites to dwarf hosts. The summer student can take a look at TNG50 halos and play with various SHMR to resolve this concern. One can also take the hosts in TNG50 and run SatGen for each of these hosts.

\begin{acknowledgments}
J.L. is grateful for the discussions with Yifei Luo, Xiaojing Lin, Yilun Ma, Risa Wechsler, Burçin Mutlu-Pakdil, Laura Hunter, Stacy Kim, and Ethan Nadler. We thank Lee Kelvin for his instructions on HSC data reduction. J.L. thanks Khalil Fong for his musical legacy. 

% HSC-SSP and HSC-LA
The Hyper Suprime-Cam (HSC) collaboration includes the astronomical communities of Japan and Taiwan, and Princeton University. The HSC instrumentation and software were developed by the National Astronomical Observatory of Japan (NAOJ), the Kavli Institute for the Physics and Mathematics of the Universe (Kavli IPMU), the University of Tokyo, the High Energy Accelerator Research Organization (KEK), the Academia Sinica Institute for Astronomy and Astrophysics in Taiwan (ASIAA), and Princeton University. Funding was contributed by the FIRST program from the Japanese Cabinet Office, the Ministry of Education, Culture, Sports, Science and Technology (MEXT), the Japan Society for the Promotion of Science (JSPS), Japan Science and Technology Agency (JST), the Toray Science Foundation, NAOJ, Kavli IPMU, KEK, ASIAA, and Princeton University. 

This paper is based on data from the Hyper Suprime-Cam Legacy Archive (HSCLA), and the data collected at the Subaru Telescope and retrieved from the HSC data archive system, which is operated by the Subaru Telescope and Astronomy Data Center (ADC) at the National Astronomical Observatory of Japan. Data analysis was in part carried out with the cooperation of the Center for Computational Astrophysics (CfCA), National Astronomical Observatory of Japan. The Subaru Telescope is honored and grateful for the opportunity to observe the Universe from Maunakea, which has cultural, historical, and natural significance in Hawaii. 

This paper makes use of software developed for the Vera C. Rubin Observatory. We thank the observatory for making their code available as free software at http://dm.lsst.org.

% DECaLS
The Legacy Surveys consist of three individual and complementary projects: the Dark Energy Camera Legacy Survey (DECaLS; Proposal ID \#2014B-0404; PIs: David Schlegel and Arjun Dey), the Beijing-Arizona Sky Survey (BASS; NOAO Prop. ID \#2015A-0801; PIs: Zhou Xu and Xiaohui Fan), and the Mayall z-band Legacy Survey (MzLS; Prop. ID \#2016A-0453; PI: Arjun Dey). DECaLS, BASS and MzLS together include data obtained, respectively, at the Blanco telescope, Cerro Tololo Inter-American Observatory, NSF’s NOIRLab; the Bok telescope, Steward Observatory, University of Arizona; and the Mayall telescope, Kitt Peak National Observatory, NOIRLab. Pipeline processing and analyses of the data were supported by NOIRLab and the Lawrence Berkeley National Laboratory (LBNL). The Legacy Surveys project is honored to be permitted to conduct astronomical research on Iolkam Du’ag (Kitt Peak), a mountain with particular significance to the Tohono O’odham Nation. The Legacy Surveys imaging of the DESI footprint is supported by the Director, Office of Science, Office of High Energy Physics of the U.S. Department of Energy under Contract No. DE-AC02-05CH1123, by the National Energy Research Scientific Computing Center, a DOE Office of Science User Facility under the same contract; and by the U.S. National Science Foundation, Division of Astronomical Sciences under Contract No. AST-0950945 to NOAO.

NOIRLab is operated by the Association of Universities for Research in Astronomy (AURA) under a cooperative agreement with the National Science Foundation. LBNL is managed by the Regents of the University of California under contract to the U.S. Department of Energy.

This project used data obtained with the Dark Energy Camera (DECam), which was constructed by the Dark Energy Survey (DES) collaboration. Funding for the DES Projects has been provided by the U.S. Department of Energy, the U.S. National Science Foundation, the Ministry of Science and Education of Spain, the Science and Technology Facilities Council of the United Kingdom, the Higher Education Funding Council for England, the National Center for Supercomputing Applications at the University of Illinois at Urbana-Champaign, the Kavli Institute of Cosmological Physics at the University of Chicago, Center for Cosmology and Astro-Particle Physics at the Ohio State University, the Mitchell Institute for Fundamental Physics and Astronomy at Texas A\&M University, Financiadora de Estudos e Projetos, Fundacao Carlos Chagas Filho de Amparo, Financiadora de Estudos e Projetos, Fundacao Carlos Chagas Filho de Amparo a Pesquisa do Estado do Rio de Janeiro, Conselho Nacional de Desenvolvimento Cientifico e Tecnologico and the Ministerio da Ciencia, Tecnologia e Inovacao, the Deutsche Forschungsgemeinschaft and the Collaborating Institutions in the Dark Energy Survey. The Collaborating Institutions are Argonne National Laboratory, the University of California at Santa Cruz, the University of Cambridge, Centro de Investigaciones Energeticas, Medioambientales y Tecnologicas-Madrid, the University of Chicago, University College London, the DES-Brazil Consortium, the University of Edinburgh, the Eidgenossische Technische Hochschule (ETH) Zurich, Fermi National Accelerator Laboratory, the University of Illinois at Urbana-Champaign, the Institut de Ciencies de l’Espai (IEEC/CSIC), the Institut de Fisica d’Altes Energies, Lawrence Berkeley National Laboratory, the Ludwig Maximilians Universitat Munchen and the associated Excellence Cluster Universe, the University of Michigan, NSF’s NOIRLab, the University of Nottingham, the Ohio State University, the University of Pennsylvania, the University of Portsmouth, SLAC National Accelerator Laboratory, Stanford University, the University of Sussex, and Texas A\&M University.

BASS is a key project of the Telescope Access Program (TAP), which has been funded by the National Astronomical Observatories of China, the Chinese Academy of Sciences (the Strategic Priority Research Program “The Emergence of Cosmological Structures” Grant \# XDB09000000), and the Special Fund for Astronomy from the Ministry of Finance. The BASS is also supported by the External Cooperation Program of Chinese Academy of Sciences (Grant \# 114A11KYSB20160057), and Chinese National Natural Science Foundation (Grant \# 12120101003, \# 11433005).

The Siena Galaxy Atlas was made possible by funding support from the U.S. Department of Energy, Office of Science, Office of High Energy Physics under Award Number DE-SC0020086 and from the National Science Foundation under grant AST-1616414.

%DESI
This research used data obtained with the Dark Energy Spectroscopic Instrument (DESI). DESI construction and operations is managed by the Lawrence Berkeley National Laboratory. This material is based upon work supported by the U.S. Department of Energy, Office of Science, Office of High-Energy Physics, under Contract No. DE–AC02–05CH11231, and by the National Energy Research Scientific Computing Center, a DOE Office of Science User Facility under the same contract. Additional support for DESI was provided by the U.S. National Science Foundation (NSF), Division of Astronomical Sciences under Contract No. AST-0950945 to the NSF’s National Optical-Infrared Astronomy Research Laboratory; the Science and Technology Facilities Council of the United Kingdom; the Gordon and Betty Moore Foundation; the Heising-Simons Foundation; the French Alternative Energies and Atomic Energy Commission (CEA); the National Council of Humanities, Science and Technology of Mexico (CONAHCYT); the Ministry of Science and Innovation of Spain (MICINN), and by the DESI Member Institutions: www.desi.lbl.gov/collaborating-institutions. The DESI collaboration is honored to be permitted to conduct scientific research on I’oligam Du’ag (Kitt Peak), a mountain with particular significance to the Tohono O’odham Nation. Any opinions, findings, and conclusions or recommendations expressed in this material are those of the author(s) and do not necessarily reflect the views of the U.S. National Science Foundation, the U.S. Department of Energy, or any of the listed funding agencies.

% Princeton clusters
The authors are pleased to acknowledge that the work reported in this paper was substantially performed using the Princeton Research Computing resources at Princeton University, a consortium of groups led by the Princeton Institute for Computational Science and Engineering (PICSciE) and the Office of Information Technology's Research Computing.

This research has made use of the SIMBAD database, operated at CDS, Strasbourg, France. The NASA/IPAC Extragalactic Database (NED) is funded by the National Aeronautics and Space Administration and operated by the California Institute of Technology. 

\end{acknowledgments}

\vspace{1em}
\facilities{Subaru (HSC), CTIO: Blanco (DECam), CFHT (MegaCam)}

\software{\href{http://www.numpy.org}{\code{NumPy}} \citep{Numpy},
          \href{https://www.astropy.org/}{\code{Astropy}} \citep{astropy}, \href{https://www.scipy.org}{\code{SciPy}} \citep{scipy}, \href{https://matplotlib.org}{\code{Matplotlib}} \citep{matplotlib},
          \href{https://artpop.readthedocs.io/en/latest/index.html}{\code{ArtPop}} \citep{artpop},
          \href{https://www.astromatic.net/software/sextractor/}{\code{SExtractor}} \citep{Bertin1996},
          \href{https://www.astromatic.net/software/swarp/}{\code{SWarp}} \citep{swarp},
          \href{https://www.astromatic.net/software/scamp/}{\code{SCAMP}} \citep{scamp},
          \href{https://www.astromatic.net/software/psfex/}{\code{PSFEx}} \citep{psfex},
          \href{https://sep.readthedocs.io/en/v1.1.x/}{\code{sep}} \citep{Barbary2016},
          \href{https://www.mpe.mpg.de/~erwin/code/imfit/}{\code{imfit}} \citep{imfit},
          \href{https://github.com/kbarbary/sfdmap}{\code{sfdmap}},
          \href{https://github.com/dr-guangtou/unagi/}{\code{unagi}},
          \href{https://github.com/shergreen/SatGen}{\code{SatGen} \citep{Jiang2021}}
          }

\bibliography{citation,software}
\bibliographystyle{aasjournal}

\newpage
\appendix

\end{CJK*}
\end{document}

%% file: hosts_cat.tex
\setlength{\tabcolsep}{4pt}
\begin{deluxetable*}{lcclcccccccc}
\tabletypesize{\small}
\tablecaption{Properties of Dwarf Hosts in This Work \label{tab:hosts}}
\tablehead{
\colhead{Name} & \colhead{$D_{\rm TRBG}$} & \colhead{$v_{\rm h}$} & \colhead{$\log L_{\rm K_s}$} & \colhead{$M_V$} & \colhead{$g - r$} & \colhead{$\log M_\star$} & \colhead{$R_{\rm vir}$} & \colhead{Area} & \colhead{$f_{\rm unmask}$} & \colhead{Data} & \colhead{Ref.} \\
\colhead{} & \colhead{(Mpc)} & \colhead{(km~s$^{-1}$)} & \colhead{($L_\odot$)} & \colhead{(mag)} & \colhead{(mag)} & \colhead{($M_\odot$)} & \colhead{(kpc)} &  \colhead{(deg$^{2}$)} & \colhead{} & \colhead{} & \colhead{}
}
\startdata
NGC5238   & $4.51\pm0.10$  & 229 & $8.02$ & $-15.0$ & $0.42$ & 7.90 & 73 & 3.923 & 94\% & B--H$_{\rm LA}$-B & \citetalias{Tully2013}; \citetalias{Karachentsev2013} \\
UGC00685  & $4.71\pm0.08$  & 156 & $8.03$ & $-14.9$ & $0.47$ & $7.94$ & 74 & 3.433 & 95\% & D--H$_{\rm LA}$--D & \citetalias{Tully2013}; \citetalias{Karachentsev2013} \\
NGC4605  & $5.55\pm0.10$  & 151 & $9.70$ & $-18.3$  & $0.59$ & 9.48 & 130 & 8.143 & 94\% & B--H,C--B & \citetalias{Tully2013}; \citetalias{Karachentsev2013} \\
NGC4707  &  $6.52\pm0.10$ & 468 & $8.25$ & $-15.7$ & $0.39$ & 8.12 & 79 & 2.553 & 95\% & B--H$_{\rm LA}$--B & \citetalias{Tully2016}; \citetalias{Karachentsev2013} \\
UGC05427 & $7.69\pm 0.10$ & 498 & 8.20 & $-15.1$ & $0.39$ & 7.89 & 72 & 1.872 & 94\% & D--H--D & \citetalias{Tully2013}; \citetalias{Karachentsev2013} \\
UGC05423 & $8.87\pm0.10$  & 348 & $8.42$ & $-15.3$ & $0.56$ & 8.25 & 83 & 1.8 &  94\% & B--H$_{\rm LA}$--B & \citetalias{Tully2013}; \citetalias{Fingerhut2010} \\
DDO046 &  $10.38\pm 0.10$ & 364 & 8.00 & $-15.5$ & $0.23$ & 7.81 & 70 & 1.309 & 80\% & B--H--B & \citetalias{Tully2013}; \citetalias{Fingerhut2010} \\
NGC4625 &  $11.75\pm 0.50$ & 606 & 9.56 & $-17.8$ & $0.56$ & 9.22 & 118 & 2.181 & 92\% & B--H--B & \citetalias{McQuinn2017}; \citetalias{Karachentsev2013} \\
\enddata
\tablecomments{The table lists the properties of the host galaxies, including the TRGB distance, heliocentric radial velocity ($v_h$), $K_s$-band luminosity, $V$-band absolute magnitude, $g-r$ color, estimated stellar mass $M_\star$, searched radius, unmasked fraction, and data sources for each host. The reference column includes the sources of distance and the $K_s$-band luminosity, respectively. The radial velocities are taken from \citetalias{Karachentsev2013}. The photometry of these dwarf hosts in the $g$ and $r$ bands is from the Siena Galaxy Atlas \citep{Moustakas2023} using the Legacy Surveys data \citep{Dey2019}, and we convert them to $V$-band using the filter conversion from \citet{DES-DR2}. The data source column lists the datasets used for satellite detection, distance measurement, and photometry, respectively. 
The letters stand for the following: B--BASS; D--DECaLS; H$_{\rm LA}$--Subaru/HSC-LA; H--Subaru/HSC; C--CFHT/MegaCam. The references stand for: 
\citetalias{Tully2013} -- \citet{Tully2013}; \citetalias{Karachentsev2013} -- \citet{Karachentsev2013}; \citetalias{Fingerhut2010} -- \citet{Fingerhut2010}; \citetalias{Tully2016} -- \citet{Tully2016}; \citetalias{McQuinn2017} -- \citet{McQuinn2017}. The $K_s$-band luminosity of UGC00685 is not from 2MASS but is converted from the $B$-band photometry and should be used with caution. UGC05423 is also known as M81-dwB \citep{Chiboucas2013}. DDO046 is also known as UGC03966.
\vspace{-2em}
}
\end{deluxetable*}

%% file: sat_cat.tex
\setlength{\tabcolsep}{2pt}
\begin{longrotatetable}
\begin{deluxetable*}{lCCcCcccCCCCCc}\label{tab:sats}
\tabletypesize{\small}
\tablecaption{SBF Distances and Photometric Properties of Satellite Candidates}
\tablewidth{0pt}
\tablehead{
    \colhead{Name} & \colhead{R.A.} & \colhead{Dec.} & 
    \colhead{$D_{\rm SBF}$} & \colhead{(S/N)$_{\rm SBF}$} & 
    \colhead{Status} & \colhead{$P_{\rm sat}$} & \colhead{$R_{\rm proj}$} & \colhead{$m_g$} & \colhead{$g-r$} & \colhead{$M_V$} & \colhead{$\log M_\star$} & \colhead{$r_{\rm eff}$} & \colhead{Completeness} \\
    \colhead{} & \colhead{(deg)} & \colhead{(deg)} &  
    \colhead{(Mpc)} & \colhead{} & \colhead{} & \colhead{} & \colhead{($R_{\rm vir}$)} & \colhead{(mag)} & \colhead{(mag)} & \colhead{(mag)} & \colhead{($M_\odot$)} & \colhead{(kpc)} & \colhead{}
}
\startdata
\multicolumn{14}{c}{NGC~5238 (4.51 Mpc)} \\ \hline 
dw1327p5141 & 201.7554 & 51.6859 & $>11.61$ & 1.0 & Rej. & 0.0 & 1.29 & 19.08 \pm 0.11 & 0.39 \pm 0.05 & -9.40 \pm 0.11 & 5.61 \pm 0.09 & 0.17 \pm 0.02 & 1.00 \\
dw1336p5108 & 204.0310 & 51.1464 & $>9.57$ & 0.8 & Rej. & 0.0 & 0.56 & 19.07 \pm 0.11 & 0.64 \pm 0.05 & -9.53 \pm 0.11 & 6.05 \pm 0.09 & 0.15 \pm 0.01 & 1.00 \\
\hline
\multicolumn{14}{c}{UGC00685 (4.81 Mpc)} \\ \hline 
dw0110p1658 & 17.6234 & 16.9814 & $>12.21$ & -0.5 & Rej. & 0.0 & 0.89 & 20.23 \pm 0.15 & 0.61 \pm 0.07 & -8.50 \pm 0.15 & 5.60 \pm 0.12 & 0.16 \pm 0.02 & 1.00 \\
dw0106p1616b & 16.5218 & 16.2809 & \nodata & \nodata & No Obs. & 0.05 & 0.56 & 20.15 \pm 0.14 & 0.37 \pm 0.06 & -8.41 \pm 0.15 & 5.18 \pm 0.12 & 0.10 \pm 0.01 & 0.95 \\
\hline
\multicolumn{14}{c}{NGC~4605 (5.55 Mpc)} \\ \hline 
dw1245p6158 & 191.4562 & 61.9695 & $4.72^{+0.56,1.09}_{-0.61,1.23}$ & 18.9 & Conf. & 1.0 & 0.58 & 18.08 \pm 0.10 & 0.61 \pm 0.04 & -10.96 \pm 0.10 & 6.57 \pm 0.08 & 0.40 \pm 0.03 & 1.00 \\
dw1247p6021 & 191.8176 & 60.3597 & $>2.58$ & 0.3 & Unconf. & 0.14 & 1.14 & 19.77 \pm 0.13 & 0.13 \pm 0.06 & -9.03 \pm 0.13 & 5.04 \pm 0.11 & 0.11 \pm 0.01 & 0.98 \\
\hline
\multicolumn{14}{c}{NGC~4707 (6.52 Mpc)} \\ \hline 
dw1246p5136 & 191.7344 & 51.6128 & $7.04^{+1.21,2.89}_{-0.84,1.49}$ & 23.8 & Conf. & 1.0 & 0.72 & 13.49 \pm 0.10 & 0.26 \pm 0.04 & -15.73 \pm 0.10 & 7.94 \pm 0.08 & 0.62 \pm 0.03 & 1.00 \\
dw1250p5056 & 192.7354 & 50.9346 & $7.81^{+1.88,4.39}_{-1.31,2.27}$ & 9.6 & Conf. & 1.0 & 0.67 & 18.88 \pm 0.11 & 0.18 \pm 0.05 & -10.30 \pm 0.11 & 5.64 \pm 0.09 & 0.13 \pm 0.01 & 1.00 \\
dw1252p5038 & 193.2005 & 50.6484 & $>9.92$ & 1.2 & Rej. & 0.0 & 1.25 & 19.77 \pm 0.13 & 0.47 \pm 0.06 & -9.55 \pm 0.13 & 5.79 \pm 0.11 & 0.15 \pm 0.02 & 0.92 \\
dw1252p5109 & 192.9984 & 51.1519 & $>12.24$ & 0.7 & Rej. & 0.0 & 0.82 & 20.55 \pm 0.16 & 0.42 \pm 0.08 & -8.74 \pm 0.16 & 5.39 \pm 0.14 & 0.13 \pm 0.02 & 0.88 \\
dw1252p5116 & 193.0036 & 51.2782 & $13.54^{+2.52,6.21}_{-1.86,3.29}$ & 4.2 & Rej. & 0.0 & 0.84 & 19.09 \pm 0.11 & 0.38 \pm 0.05 & -10.18 \pm 0.11 & 5.90 \pm 0.09 & 0.21 \pm 0.02 & 1.00 \\
\hline
\multicolumn{14}{c}{UGC05427 (7.69 Mpc)} \\ \hline 
dw1006p2852 & 151.5974 & 28.8717 & $>12.63$ & 0.2 & Rej. & 0.0 & 1.15 & 19.62 \pm 0.13 & 0.56 \pm 0.05 & -10.10 \pm 0.13 & 6.15 \pm 0.10 & 0.31 \pm 0.03 & 1.00 \\
dw1003p2935 & 150.8486 & 29.5926 & $12.14^{+2.89,7.00}_{-2.10,3.61}$ & 5.9 & Unconf. & 0.09 & 0.67 & 20.18 \pm 0.14 & 0.22 \pm 0.06 & -9.37 \pm 0.15 & 5.32 \pm 0.12 & 0.13 \pm 0.02 & 0.97 \\
dw1003p3004 & 150.8554 & 30.0721 & \nodata & \nodata & No Obs. & 0.67 & 1.41 & 19.04 \pm 0.11 & 0.39 \pm 0.05 & -10.59 \pm 0.11 & 6.08 \pm 0.09 & 0.30 \pm 0.03 & 1.00 \\
\hline
\multicolumn{14}{c}{UGC05423 (8.87 Mpc)} \\ \hline 
dw1008p7038b & 152.2005 & 70.6456 & $10.63^{+1.49,3.35}_{-1.16,2.15}$ & 36.8 & Conf. & 1.0 & 0.73 & 17.78 \pm 0.20 & 0.33 \pm 0.12 & -12.08 \pm 0.21 & 6.61 \pm 0.22 & 0.45 \pm 0.03 & 1.00 \\
dw1009p7032 & 152.3942 & 70.5489 & $9.13^{+0.76,1.59}_{-0.66,1.26}$ & 14.4 & Conf. & 1.0 & 0.72 & 17.93 \pm 0.10 & 0.32 \pm 0.04 & -11.98 \pm 0.10 & 6.52 \pm 0.08 & 0.55 \pm 0.04 & 1.00 \\
dw1008p7038 & 152.2106 & 70.6339 & $14.62^{+1.54,3.30}_{-1.42,2.88}$ & 51.5 & Rej. & 0.0 & 0.72 & 16.11 \pm 0.09 & 0.60 \pm 0.04 & -13.94 \pm 0.09 & 7.75 \pm 0.07 & 0.80 \pm 0.04 & 1.00 \\
dw1003p7040 & 150.9903 & 70.6672 & $>12.12$ & 1.6 & Rej. & 0.0 & 0.61 & 21.49 \pm 0.22 & 0.56 \pm 0.13 & -8.54 \pm 0.23 & 5.53 \pm 0.24 & 0.16 \pm 0.03 & 0.61 \\
dw1009p7030 & 152.3846 & 70.5053 & $>17.74$ & -0.8 & Rej. & 0.0 & 0.68 & 21.13 \pm 0.19 & 0.54 \pm 0.10 & -8.89 \pm 0.20 & 5.64 \pm 0.19 & 0.32 \pm 0.06 & 0.43 \\
dw1008p7058 & 152.1400 & 70.9753 & \nodata & \nodata & No Obs. & 0.58 & 1.23 & 19.81 \pm 0.13 & 0.39 \pm 0.06 & -10.14 \pm 0.14 & 5.90 \pm 0.11 & 0.44 \pm 0.05 & 0.83 \\
\hline
\multicolumn{14}{c}{DDO046 (10.38 Mpc)} \\ \hline 
dw0741p4005 & 115.3924 & 40.0843 & $10.28^{+2.57,6.03}_{-1.96,3.42}$ & 5.1 & Conf. & 1.0 & 0.10 & 18.46 \pm 0.10 & 0.17 \pm 0.04 & -11.72 \pm 0.10 & 6.18 \pm 0.08 & 0.28 \pm 0.02 & 1.00 \\
\hline
\multicolumn{14}{c}{NGC~4625 (11.75 Mpc)} \\ \hline 
dw1241p4103 & 190.2539 & 41.0530 & $18.48^{+2.10,4.32}_{-1.92,3.78}$ & 7.8 & Rej. & 0.0 & 0.48 & 18.84 \pm 0.11 & 0.52 \pm 0.04 & -11.79 \pm 0.11 & 6.77 \pm 0.09 & 0.41 \pm 0.04 & 0.93 \\
dw1242p4115 & 190.5470 & 41.2525 & $7.96^{+0.76,1.47}_{-0.77,1.51}$ & 29.7 & Rej. & 0.0 & 0.11 & 18.23 \pm 0.10 & 0.57 \pm 0.04 & -12.42 \pm 0.10 & 7.10 \pm 0.08 & 0.73 \pm 0.06 & 1.00 \\
dw1243p4056 & 190.9151 & 40.9409 & $>16.92$ & 0.1 & Rej. & 0.0 & 0.82 & 22.20 \pm 0.29 & 0.47 \pm 0.21 & -8.40 \pm 0.31 & 5.33 \pm 0.38 & 0.18 \pm 0.05 & 0.44 \\
dw1244p4126 & 191.2270 & 41.4403 & $19.81^{+3.76,8.85}_{-2.82,5.16}$ & 4.1 & Rej. & 0.0 & 1.03 & 19.36 \pm 0.12 & 0.58 \pm 0.05 & -11.30 \pm 0.12 & 6.67 \pm 0.10 & 0.32 \pm 0.03 & 1.00 \\
dw1244p4118 & 191.0675 & 41.2999 & $19.49^{+8.30,inf}_{-4.61,7.97}$ & 2.1 & Unconf. & 0.61 & 0.78 & 21.19 \pm 0.19 & 0.54 \pm 0.11 & -9.45 \pm 0.20 & 5.86 \pm 0.20 & 0.24 \pm 0.05 & 1.00 \\
dw1240p4205 & 190.0652 & 42.0986 & \nodata & \nodata & No Obs. & 0.48 & 1.53 & 21.68 \pm 0.24 & 0.40 \pm 0.15 & -8.89 \pm 0.25 & 5.41 \pm 0.27 & 0.18 \pm 0.04 & 0.75 \\
dw1238p4043 & 189.6391 & 40.7237 & \nodata & \nodata & No Obs. & 0.23 & 1.45 & 19.66 \pm 0.13 & 0.43 \pm 0.05 & -10.92 \pm 0.13 & 6.28 \pm 0.10 & 0.21 \pm 0.02 & 1.00 \\
dw1239p4119 & 189.9575 & 41.3297 & \nodata & \nodata & No Obs. & 0.54 & 0.68 & 21.74 \pm 0.24 & 0.38 \pm 0.15 & -8.81 \pm 0.26 & 5.36 \pm 0.28 & 0.16 \pm 0.04 & 0.75 \\
dw1241p4128 & 190.2598 & 41.4767 & \nodata & \nodata & No Obs. & 0.14 & 0.45 & 21.49 \pm 0.22 & 0.26 \pm 0.13 & -9.01 \pm 0.23 & 5.25 \pm 0.24 & 0.13 \pm 0.03 & 0.43 \\
\hline
\enddata
\tablecomments{roperties of satellite candidates, including name, R.A., Dec., the signal-to-noise ratio of SBF measurement, confirmation status, satellite probability ($P_{\rm sat}$), projected distance from the host in unit of host virial radius, $g$-band apparent magnitude, $g-r$ color, $V$-band absolute magnitude ($M_V$), estimated stellar mass ($\log M_\star$), and half-light radius ($r_{\rm eff}$). The last three columns are estimated assuming the candidates are at the distance of the host, and are only meaningful for the confirmed ones. 
}
\end{deluxetable*}
\end{longrotatetable}

%% file: literature_hosts.tex
\setlength{\tabcolsep}{4pt}
\begin{longrotatetable}
\begin{deluxetable*}{lLLCCCCCCCcp{4cm}}
\tabletypesize{\footnotesize}
\tablecaption{Classical Satellites of Dwarf Galaxies from the Literature \label{tab:literature}}
\tablewidth{0pt}
\tablehead{
\colhead{Name} & \colhead{R.A.} & \colhead{Decl.} & \colhead{Distance} & \colhead{$M_V$} & 
\colhead{$\log M_{\star}$} & \colhead{$r_{\rm eff}$} & \colhead{$\Theta_1$} & \colhead{$\Theta_5$} & \colhead{$\Theta_j$} & \colhead{Conf.} & \colhead{Ref.} \\
\colhead{} & \colhead{(deg)} & \colhead{(deg)} & \colhead{(Mpc)} & \colhead{(mag)} & 
\colhead{$M_\odot$} & \colhead{(kpc)} & \colhead{} & \colhead{} & \colhead{} & \colhead{} & \colhead{}
}
\startdata
LMC & 80.8942 & -69.7561 & 0.05 & -18.1 & 9.3 & \nodata & 3.5 & 3.6 & 1.7 & \nodata & \citet{Skibba2012LMC} \\
\hspace{1em}SMC & 13.1580 & -72.8003 & 0.06 & -16.8 & 8.5 & 1.08 & \nodata & \nodata & \nodata & Y & \citet{Pace2024} \\
\hline
NGC~6822 & 296.2404 & -14.8031 & 0.46 & -15.2 & 8.1 & \nodata & 0.5 & 0.6 & 1.7 & \nodata & \citet{Zhang2021_NGC6822} \\
\hspace{1em} No Satellites \\
\hline
M33 & 23.4617 & 30.6603 & 0.84 & -18.8 & 9.7 & \nodata & 1.5 & 1.5 & 1.7 & \nodata & \citet{Corbelli2014} \\
\hspace{1em} No Satellites \\
\hline
NGC~3109 & 150.7769 & -26.1549 & 1.34 & -14.9 & 8.2 & \nodata & -0.3 & -0.1 & -1.2 & \nodata & \citetalias{McConnachie2012} \\
\hspace{1em}Antlia & 151.0171 & -27.3311 & 1.35 & -10.4 & 6.4 & 0.46 & \nodata & \nodata & \nodata & Y & \citetalias{McConnachie2012} \\
\hspace{1em}Antlia B & 147.2337 & -25.9900 & 1.29 & -9.7 & 6.2 & 0.27 & \nodata & \nodata & \nodata & Y & \citet{Sand2015} \\
\hline
NGC~300 & 13.7227 & -37.6842 & 2.00 & -18.4 & 9.2 & \nodata & 0.1 & 0.2 & 0.2 & \nodata & \nodata \\
\hspace{1em}Sculptor C & 14.4675 & -35.8523 & 2.04 & -9.1 & 5.7 & 0.36 & \nodata & \nodata & \nodata & Y & \nodata \\
\hline
NGC~55 & 3.7208 & -39.1966 & 2.34 & -18.5 & 9.7 & \nodata & 0.0 & 0.2 & 0.2 & \nodata & \citet{Kudritzki2016} \\
\hspace{1em}NGC55-dw1 & 3.8740 & -38.4190 & 2.20 & -8.0 & 5.2 & 2.20 & \nodata & \nodata & \nodata & Y & \citet{McNanna2024} \\
\hline
NGC~4214 & 183.9121 & 36.3275 & 3.04 & -17.5 & 8.8 & \nodata & 0.6 & 0.8 & -0.6 & \nodata & \citet{Dalcanton2009} \\
\hspace{1em}MADCASH-2 & 182.5281 & 35.4429 & 3.00 & -9.2 & 5.8 & 0.13 & \nodata & \nodata & \nodata & Y & \citet{Carlin2021} \\
\hspace{1em}DDO113 & 183.7412 & 36.2189 & 2.95 & -12.2 & 7.1 & 0.62 & \nodata & \nodata & \nodata & Y & \citet{Garling2020} \\
\hline
NGC~2403 & 114.2117 & 65.6022 & 3.19 & -19.1 & 9.7 & \nodata & 0.2 & 0.6 & 1.9 & \nodata & \citetalias{Karachentsev2013} \\
\hspace{1em}MADCASH-1 & 114.2140 & 65.6010 & 3.41 & -7.8 & 5.3 & 0.18 & \nodata & \nodata & \nodata & Y & \citet{Carlin2016,Carlin2024} \\
\hspace{1em}DDO044 & 113.5479 & 66.8797 & 2.96 & -12.9 & 7.3 & 0.74 & \nodata & \nodata & \nodata & Y & \citet{Karachentsev1999DDO044,Carlin2019} \\
\hline
NGC~3077 & 150.8375 & 68.7339 & 3.85 & -17.8 & 9.4 & \nodata & 2.3 & 2.4 & 1.9 & \nodata & \citet{Garling2021} \\
\hspace{1em} No Satellites \\
\hline
IC~2574 & 157.0933 & 68.4161 & 3.93 & -17.1 & 8.8 & \nodata & 1.5 & 1.7 & 1.9 & \nodata & \citet{Garling2021} \\
\hspace{1em} No Satellites \\
\hline
NGC~4449 & 187.0458 & 44.0936 & 4.02 & -18.4 & 9.5 & \nodata & 0.8 & 0.8 & 1.4 & \nodata & \citet{Sabbi2018} \\
\hspace{1em}LV J1228+4358 & 187.1867 & 43.9683 & 4.07 & -13.0 & 7.0 & 1.30 & \nodata & \nodata & \nodata & Y & \citet{Martinez-Delgado2012,Garling2021} \\
\hline
NGC~4236 & 184.1804 & 69.4656 & 4.44 & -18.2 & 9.4 & \nodata & -0.2 & 0.0 & 0.0 & \nodata & \citet{Garling2021} \\
\hspace{1em} No Satellites \\
\hline
NGC~5023 & 198.0496 & 44.0386 & 6.05 & -16.5 & 8.8 & \nodata & -0.9 & -0.3 & -1.1 & \nodata & \citetalias{Karachentsev2013} \\
\hspace{1em}dw1310+4358 & 197.7499 & 43.9802 & \nodata & -8.0 & 5.7 & 0.12 & \nodata & \nodata & \nodata & N & \citet{Carlsten2020} \\
\hspace{1em}dw1314+4420 & 198.6433 & 44.3342 & \nodata & -7.0 & 5.0 & 0.10 & \nodata & \nodata & \nodata & N & \citet{Carlsten2020} \\
\hline
NGC~1156 & 44.9267 & 25.2375 & 6.87 & -17.4 & 9.0 & \nodata & -1.5 & -1.2 & -1.3 & \nodata & \citetalias{Karachentsev2013} \\
\hspace{1em}NGC1156-dw1 & 45.0758 & 25.2489 & \nodata & -10.5 & 6.1 & \nodata & \nodata & \nodata & \nodata & N & \citetalias{Karachentsev2013} \\
\hspace{1em}NGC1156-dw2 & 45.1138 & 25.3051 & \nodata & -10.1 & 6.3 & \nodata & \nodata & \nodata & \nodata & N & \citetalias{Karachentsev2013} \\
\hspace{1em}dw0301+2446 & 45.3842 & 24.7832 & \nodata & -10.6 & 6.5 & \nodata & \nodata & \nodata & \nodata & N & \citet{Carlsten2020} \\
\hspace{1em}AGES J030039+254656 & 45.1625 & 25.7822 & \nodata & \nodata & \nodata & \nodata & \nodata & \nodata & \nodata & N & \citet{Minchin2010} \\
\hline
NGC~45 & 3.5162 & -23.1822 & 6.64 & -20.0 & 9.1 & \nodata & -1.0 & -0.9 & 0.2 & \nodata & \citet{Muller2020} \\
\hspace{1em} No Satellites \\
\hline
NGC~672 & 26.9717 & 27.4336 & 7.18 & -18.2 & 9.5 & \nodata & 0.2 & 0.3 & 0.2 & \nodata & \citetalias{Karachentsev2013} \\
\hspace{1em}IC1727 & 26.8745 & 27.3333 & 7.45 & -17.3 & 9.3 & 3.30 & \nodata & \nodata & \nodata & Y & \citetalias{Karachentsev2013} \\
\hspace{1em}NGC672-dwA & 26.8304 & 27.2548 & \nodata & -9.4 & 5.5 & 0.20 & \nodata & \nodata & \nodata & N & \citet{Davis2024} \\
\hline
NGC~24 & 2.4850 & -24.9633 & 7.31 & -20.3 & 9.2 & \nodata & 0.3 & 0.6 & 0.1 & \nodata & \citetalias{Karachentsev2013} \\
\hspace{1em}dw0009-25 & 2.4075 & -25.0492 & \nodata & -8.3 & 5.9 & 0.28 & \nodata & \nodata & \nodata & N & \citet{Muller2020} \\
\hspace{1em}dw0010-25 & 2.6592 & -25.3349 & \nodata & -11.6 & 6.7 & 1.05 & \nodata & \nodata & \nodata & N & \citet{Muller2020} \\
\hline
NGC~925 & 36.8167 & 33.5781 & 9.55 & -19.8 & 9.9 & \nodata & 1.0 & 1.1 & 0.5 & \nodata & \citetalias{Karachentsev2013} \\
\hspace{1em}d0226+3325 & 36.7200 & 33.4269 & \nodata & -12.0 & 7.0 & 1.50 & \nodata & \nodata & \nodata & N & \citet{PWE98,Davis2024} \\
\hline
NGC~4592 & 189.8279 & -0.5319 & 9.02 & -18.0 & 9.0 & \nodata & -0.4 & 0.1 & 1.2 & \nodata & \citet{Kim2020} \\
\hspace{1em}NGC4592-dw15 & 189.7053 & -0.5978 & 9.02 & -10.8 & 6.1 & 0.23 & \nodata & \nodata & \nodata & Y & \citet{Kim2022,CarlstenELVES2022} \\
\hspace{1em}NGC4592-dw16 & 189.7608 & -0.6650 & 9.02 & -13.2 & 7.2 & 0.74 & \nodata & \nodata & \nodata & Y & \citet{Kim2022,CarlstenELVES2022} \\
\hline
M96-DF6 & 161.7212 & 12.7426 & 10.20 & -13.4 & 7.6 & \nodata & 2.0 & 2.0 & 2.0 & \nodata & \citet{Cohen2018} \\
\hspace{1em}dw1046+1244& 161.6915 & 12.7499 & \nodata & -8.1 & 5.4 & 0.21 & \nodata & \nodata & \nodata & N & \citet{Muller2023} \\
\hline
\enddata
\tablecomments{This table presents a compilation of classical satellite candidates ($M_\star > 10^{5}\ M_\odot$) associated with dwarf galaxies ($M_\star <10^{10}\ M_\odot$) in the Local Volume ($<12\ \rm Mpc$) from the literature. For each system, the host galaxy is listed first, followed by its satellite candidates. We include key properties for both hosts and satellite candidates: Right Ascension and Declination (R.A. and Decl.), distance, $V$-band absolute magnitude ($M_V$), estimated stellar mass $\log M_\star$, and physical effective radius ($r_{\rm eff}$). For host galaxies, we also provide the tidal indices ($\Theta_1$, $\Theta_5$, $\Theta_j$) from \citet{Karachentsev2013}, which characterize their environment. Satellites that are confirmed with direct distances or radial velocities are marked as ``Y'' in the Conf. column. References for each object are provided in the final column. See Appendix \ref{sec:literature} for detailed descriptions of individual systems. }
\end{deluxetable*}
\end{longrotatetable}